\documentclass[fleqn,10pt]{wlscirep}

\usepackage{amsthm,amsmath}
\usepackage[utf8]{inputenc} %unicode support
\usepackage[T1]{fontenc}
\usepackage{url}
\usepackage[capitalise]{cleveref}
\usepackage{bm}
\usepackage{adjustbox}
\usepackage{algorithm2e}

\usepackage{graphicx}
\usepackage{caption}
\usepackage{subcaption}
\graphicspath{{Figures/}}
\usepackage{float}

\usepackage{pdfpages}

\usepackage{xcite}

\usepackage{xcolor,etoolbox}
\makeatletter 
\pretocmd\@bibitem{\color{black}\csname keycolor#1\endcsname}{}{\fail}
\newcommand\citecolor[1]{\@namedef{keycolor#1}{\color{red}}}
\makeatother
\title{A new set of cluster driven composite development indicators}

\author[1,*,+]{Anshul Verma}
\author[1,+]{Orazio Angelini}
\author[1,2,3,+]{Tiziana Di Matteo}
\affil[1]{Department of Mathematics, King's College London, London}
\affil[2]{Department of Computer Science, University College London, London}
\affil[3]{Complexity Science Hub Vienna, Vienna}

\affil[*]{anshul.verma@kcl.ac.uk}

\affil[+]{Equal contributors}
\begin{abstract}
Composite development indicators used in policy making often subjectively aggregate a restricted set of indicators. We show, using dimensionality reduction techniques, including Principal Component Analysis (PCA) and for the first time information filtering and hierarchical clustering, that these composite indicators miss key information on the relationship between different indicators. In particular, the grouping of indicators via topics is not reflected in the data at a global and local level. We overcome these issues by using the clustering of indicators to build a new set of cluster driven composite development indicators that are objective, data driven, comparable between countries, and retain interpretabilty. We discuss their consequences on informing policy makers about country development, comparing them with the top PageRank indicators as a benchmark. Finally, we demonstrate that our new set of composite development indicators outperforms the benchmark on a dataset reconstruction task.
\end{abstract}

\maketitle

\begin{document}

%%%%%%%%%%%%%%%%%%%%%%%%%%%%%%%%%%%%%%%%%%%%%%
%%                                          %%
%% The keywords begin here                  %%
%%                                          %%
%% Put each keyword in separate \kwd{}.     %%
%%                                          %%
%%%%%%%%%%%%%%%%%%%%%%%%%%%%%%%%%%%%%%%%%%%%%%

%\begin{keyword}
%\kwd{Development economics}
%\kwd{Composite indicators}
%\kwd{Information filtering}
%\kwd{Clustering}
%\kwd{World Development Indicators}
%\end{keyword}

\section{Introduction}
Economic indicators are vital in understanding and tracking the macroeconomic state and development of a country \cite{Stock1989}, informing government policy makers about the health of the economy and also for citizens to evaluate and assess any improvement in their life \cite{Mugge2016}. However, with the ever expanding number of different indicators and digital records of this data, it becomes difficult to interpret the high dimensional data as a whole, spot overall trends and see how different indicators are related to each other. Often qualitative or obscure factors are used to explain development such as the need to have a good education and healthy citizens.    

Additionally, it is not agreed what factors affect development \cite{Ricardo1891,Leontief1956,Bowen1986,Aghion1990,Heckscher1991,Kremer1993,Krueger2001,Egert2009,Aghion2010}, and so arbitrarily chosen indicators are often used, %\todo{this should be reformulated. ignoring what information? the one in indicators that are excluded?}
ignoring specific information by excluding other indicators. In some cases, a more educated assumption is made by taking only indicators of relevance e.g. those relating to infrastructure. Even in these cases, different classes of indicators are treated separately to each other. Links with other classes of indicators e.g. poverty and infrastructure \cite{United1997} are disregarded.
%and with less a-priori assumptions about the structure of their interdependence.
This is especially relevant when one combines them in some way into composite indicators\cite{Salzman2003}, which aim to describe several development indicators with just one composite version. These can range from more general indicators such as the Human Development Index (HDI) \cite{Sagar1998}, which is used to measure the progress in life expectancy, education and Gross National Income per capita (GNI)\cite{Todaro2015}, to more specific indicators such as the Global Connectivity Index (GCI) \cite{GCI}. Composite indicators are also often used to summarise the state of a country relating to the specific objective of combining the chosen set of indicators e.g. GCI is used to track the extent of digital infrastructure of a country, whilst HDI is used to track overall human development. In literature they have been used, for example, to relate cancer rates to development \cite{Bray2012} or to produce a global rank of a country's competitiveness. %\cite{}%

Whilst aggregating indicators into composite ones seems to be a good solution to the problem of summarising information from many different indicators, we propose that the high number of possible ways to combine them calls for the developement of guiding principles on how this should be achieved. Moreover, some indicators are calculated differently for different regions \cite{Huggins2003}, making comparisons based on them much more difficult. By knowing how indicators are inter-related to each other, we will be able to understand in a data-driven, objective way which indicators are most important in characterising a country and how they should be combined to produce economically meaningful composite indicators. 

Dimensionality reduction can help here by providing a smaller but faithful version of the relationship between the vast number of available development indicators \cite{Maaten2009}. This paper proposes to study these relationships in an unbiased way, using these relationships as a basis to propose a new set of composite development indicators. Differently to previous work, we make no subjective restriction on the type of indicators we study, drawing from a large range of scope of indicators to study the relationship between the indicators emerging from the data itself. We test whether we can indeed separate the indicators into different pre defined groups based on the different factors proposed that affect development using PCA (Principal Component Analysis) and Random Matrix Theory (RMT) \cite{Bun2017}. We find that the broad topic category they are assigned, e.g. health vs economic vs infrastructure, is not necessarily the best way to aggregate them. We also employ hierarchical clustering algorithms for the first time to analyse the structure of indicators rather than countries, finding that the indicator clusters are a mixture of topics but still retain an economic interpretation. We use these results to overcome traditional problems faced in making composite indicators such as how/what indicators to aggregate to derive a new set of objective, data driven, interpretable and country comparable composite indicators. Leveraging on these composite development indicators, we observe useful observations for policy makers, such as the ability of mobile phone adoption to be able to distinguish between underdeveloped countries. Next, we provide a new application of network filtering to find subsets of highly influential indicators based on PageRank \cite{Page1999}. Finally, we compare the performance of our composite indicators to a random benchmark, a subset of influential indicators and PCA, concluding that our proposed composite indicators outperform the others.

%With a specific view on economic growth, a number of different factors have been proposed to influence economic growth. For example, infrastructure investment is seen as a good candidate \cite{} as the presence of a good road network, airports, trains etc all facilitate growth by enhancing productivity \cite{}. However, at the same time one can also argue that health care expenditure can also increase growth since a healthier population is more productive \cite{Mushkin1962}. One can also make the same argument for education. We see here that we can offer competing factors that can increase economic growth 
In the context of this problem, dimensionality reduction has been applied in \cite{Cristelli2018}, where Principal Component Analysis (PCA) was used on a set of restricted indicators to form infrastructure composite indicator. This indicator was then used to examine the direction of the causal relationship between infrastructure capability and economic growth. The authors of \cite{Lai2003} compare the pre-set weights of the HDI to that derived from a PCA. Hierarchical clustering has been applied to analyse clusters of countries such as in \cite{Nardo2005,Castellacci2011}. However, in either cases either a restricted set of indicators are used or the focus of the work is on countries, rather than the analysis and development of new indicators themselves. The problem of forming composite indicators using PCA and dimensionality reduction is emphasised in  \cite{Mazziotta2019}, where it has been shown this approach could overlook important information. In particular for PCA, this includes ignoring information from components other than the first and difficult to interpret weights.

Network filtering techniques \cite{Mantegna1999,Tumminello2005} and their related hierarchical clustering algorithms \cite{Anderberg2014,Song2012,Musmeci2015} have also proved to be useful when analysing data, with wide ranging applications from finance to biology \cite{Musmeci2015,Song2012,Sneath1957}. Network filtering techniques view a similarity matrix as a network, each node being a feature and each link having a weight with the respective non-zero correlation. Within this framework, removing noisy entries in the correlation matrix can be translated into finding a sparse version of the similarity network. These techniques aim to extract the backbone of the structure between generic features by enforcing sparsity in a specific way to the particular technique. The induced sparsity of the network helps make hidden structures more visible. One successful example of this is the Minimum Spanning Tree (MST) \cite{Graham1985,Mantegna1999}, which imposes that the correlation matrix is a tree that maximises the total weight of links, and has been applied in a diverse number of fields from electricity networks to taxonomy \cite{Sneath1957,Graham1985}. A generalisation which includes the possibility of loops is the Planar Maximally Filtered Graph (PMFG) \cite{Aste2005,Tumminello2005}, which instead imposes a weaker constraint that the network is planar i.e. it can be embedded on a sphere without any links crossing. Hierarchical algorithms are also highly related and aim to group features with similar properties into clusters that organised in a hierarchical fashion in the form of a dendrogram. An example of this is the Directed Bubble Hierarchical Tree (DBHT) algorithm that is based on the PMFG, having been used for finance \cite{Musmeci2015,Musmeci2015a} and in gene expression data \cite{Song2012}. In particular, the DBHT algorithm has also been shown to outperform other hierarchical clustering algorithms.        
%We aim to answer the question of whether a set of global World Development Indicators (WDI) can be shown to have an interpretable, global structure that can summarise the relationship between indicators. First we use PCA, finding a novel result that groups of indicators are linked, but do not reflect well the general topic of indicators . We then present a new analysis based on the network filtering technique PMFG that identifies key influential indicators. We then use an objective hierarchical clustering algorithm called Directed Bubble Hierarchical Tree (DBHT), finding that the empirical clustering is very different compared to the preassigned topic label. Discussing how we can use this to build an objective, data driven and interpretable set of composite indicators that can give an idea of the current state of a country, we give some interesting insights into the development of countries. Finally, we compare the performance of our composite indicators to a random benchmark and a set of high, network influential indicators, showing that our composite indicators can outperform them

This paper is organised as follows. The second section is a description of the dataset and how we amalgamate topics together. In the third section we apply a PCA analysis to our dataset, which we use to show the difference between the structure of the empirical correlation matrix and the preassigned topics. We then find the clustering using the DBHT algorithm in \cref{DBHT_interpret}. Developing the clustering results further to form a novel set of composite indicators in \cref{CompInds_DBHT}, we observe some interesting features of our composite indicators in \cref{CountryDevelopment}. For \cref{Influential_PMFG} we apply the PMFG to the empirical correlation matrix in order to derive some influential indicators via PageRank. In \cref{PerformanceComparison} we compare the performance of our composite indicators with a random benchmark and top indicators taken from the PageRank. Finally, we discuss the dynamic stability of our results in \cref{DynamicalAnalysis} and draw some conclusions in the final section.    

\section{WDI Dataset} \label{WDI_Data}
The World Development Indicator (WDI) dataset is a vast collection of various yearly development indicators for $C=218$ countries (where $C$ is the number of countries) and are taken from official, internationally recognised agencies \cite{World2018}. Note that we have applied the imputation scheme the distribution regularisation procedure detailed in the Supplementary Information \cref{Imputation} (to treat missing data) and in \cref{Distribution_regularisation} (to standardise and normalise the indicators) respectively. Note that we have also checked that the amount of missing data does not change the results significantly - see Supplementary Information \cref{Imputation} for more information. We shall use a total of $T=19$ years, where $t=1,...,19$ represents the years from $1998$ to $2016$ . The number of indicators contained within the dataset is $N=1574$, and the objectives for collecting these indicators ranges from well known economic data such as Gross Domestic Product (GDP), education data such as the literacy rate and population health such as infant mortality rate. Hence both the large number of indicators and the diverse range of granularity and objectives makes this dataset a perfect candidate in order to study the relationships between different classes of indicators and to infer and derive conclusions that hold globally. Note that we also remove highly correlated indicators with a correlation coefficient of $0.95$ since some indicators can be trivially related together e.g. percentage of population that are males and the same but for females, which would bias the results. This reduces the number of indicators $N$ to $1448$. 

\subsection{Amalgamating Topics} \label{AmalgamatingTopics}
\begin{figure}
\centering
\includegraphics[width=0.9\textwidth]{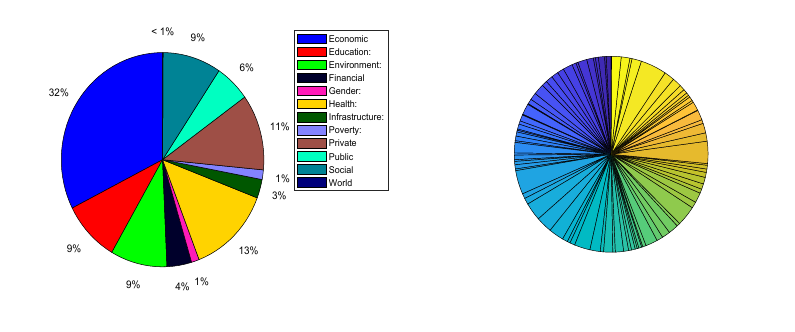}
\label{fig:pie_chart}
\caption{ (Left panel) a) Pie chart of the distribution of the indicators using the classification from the WDI dataset. (Right panel) b) The same but with the aggregated classification. The legend of the bottom chart indicates the names of the derived aggregated classification of the topics of the indicators.}
\end{figure}

The indicators are also divided into $94$ different topics that include different classes of economic indicators such as Economic Policy and Debts: National Accounts: Growth Rates, which measures growth rates of agriculture, industry, manufacturing and services sectors, and Education: Participation, which measures participation rates across gender, age groups in various levels of education. We show the distribution of all such topics using this classification in \cref{fig:pie_chart}a. We can see that most of the groups of indicators make up a very small fraction of the indicators, which would mean that any averaged statistic across within each group would be subject to significant noise. To counteract this, we aggregate the topics for each classification based on their root objective e.g. Education: Participation and Education: Efficiency are both classes of indicators relating to education and hence we combine these two groups into one group Education, similarly Health: Nutrition and Health: Disease Prevention are combined into Health. Applying this procedure to the entire dataset produces $g=1,...,G=12$ different topics for the indicators which we indicate in \cref{fig:pie_chart}b. We can see that each topic has a larger number of indicators, which will increase the statistical reliability of any conclusions drawn from the data.

%\begin{figure}
%\centering
%\begin{subfigure}{0.475\textwidth}
%\centering
%\includegraphics[width=\textwidth]{Figures/pie_chart_remove.eps}
%\caption{WDI classification}
%\label{fig:pie_chart}
%\end{subfigure}
%\\
%\begin{subfigure}{0.475\textwidth}
%\centering
%\includegraphics[width=\textwidth]{Figures/pie_chart_short.eps}
%\caption{Aggregated Classification}
%\label{fig:pie_chart_short}
%\end{subfigure}
%\caption{(Left panel) \subref{fig:pie_chart} Pie chart of the distribution of the indicators using the classification from the WDI dataset. (Right panel) \subref{fig:pie_chart_short} The same but with the aggregated classification. The legend of the bottom chart indicates the names of the derived aggregated classification of the topics of the indicators.}
%\end{figure}

\subsection{Data Structure} \label{DataStructure}
We here start exploring the correlation structure across years as a measure to quantify the relationship between indicators. We aggregate values across years in order to average out correlations that might hold only for specific periods or groups of countries. This also helps to reduce the noise in the correlation matrix, since one-year matrices would be too shallow to reliably obtain correlation estimates. %\todo{I'm not sure this is the best way to say this or the right place, but it has to be said somewhere. <ORAZIO:> agree it has to be said. Placement, I'd like to re-proofread it entirely to see if it fits best elsewhere} \todo{mention the $q=N/T$ value? <ORAZIO:> possibly, since ours is very high and this affects the whole analysis}.

With this mind, we organise the data matrix $\bm{X}$ as follows. It consists of $C$ matrices of size $T\times N$ matrices stacked vertically, with each cross sectional block representing the data for one specific country $c$ and each column reporting the data for indicator $i=1,...,1448$. For each cross section, the entries in the first row and $i$ column are the values of the indicator $i$ for $t=1$ and the last row are the same but for $t=T$. In order to discard spurious correlations in the data, we remove trends by taking the first difference, that is for each block of data we calculate
\begin{equation}
%X^{\text{diff},c}_{\tilde{y},i}=X^{c}_{y+1,i}-X^{c}_{y,i} \ ,
\Delta \bm{X}(\tilde{t},c,i)=\bm{X}(t+1,c,i)-\bm{X}(t,c,i)
\end{equation}
with $\bm{X}(t,c,i)$ is the value of indicator $i$ for country $c$ at year $t$. $\bm{X}(t+1,c,i)$ is similar but with $t+1$. $\Delta \bm{X}(\tilde{t},c,i)$ represents the first difference between $\bm{X}(t+1,c,i)$ and $\bm{X}(t,c,i)$, with $\tilde{t}$ running from $1,...,T-1=18$. Every $\Delta \bm{X}(\cdot,c,\cdot)$ has $T-1$ rows and $N$ columns. Stacking each of these vertically forms the $Y=3924\times N$ matrix $\Delta \bm{X}$, which now contains all the differenced values for all countries and all time steps.   

To encode the relationship between the indicators we use the empirical Pearson correlation matrix $\bm{E}$, which can be calculated from a zero mean, standardised $\Delta \bm{X}$ as
\begin{equation}
%E_{ij}=\frac{1}{C\tilde{Y}}\sum_{c,c'=1}^{C}\sum_{\tilde{y},\tilde{y}'=1}^{\tilde{Y}}X^{\text{diff},c}_{\tilde{y},i}X^{\text{diff},c'}_{\tilde{y}',j} \ .
\bm{E}=\frac{1}{C(T-1)}\left(\Delta \bm{X}\right)^{\dagger} \Delta \bm{X} \ , \label{CorrMatrix}
\end{equation}
where $\dagger$ represents the transpose. Therefore, we aim to understand the multivariate dependence between development indicators through analysing the main driving factors of the structure of $\mathbf{E}$. However, using the raw correlation matrix would be unwise due to its large size ($1448$ by $1448$) and noise present in the system, potentially leaving a certain amount of redundant information in $\mathbf{E}$. As mentioned earlier, we can distill the information given in $\mathbf{E}$ to a smaller version using dimensionality reduction, which should also have the added benefit of making it easier to interpret the structure of $\mathbf{E}$.

\section{PCA analysis} \label{PCA}
%Start off explaining PCA. Why do we use it? Give results on eigenvalue distribution - focus on how bulk indicates that we can't cutoff at arbitrary level. Give results on eigenvectors and explain how this means that there is a mix between indicators defined by different topics - can't analyse them in isolation. 
Within the class of dimensionality reduction methods, PCA is a popular and easy to apply technique used on correlation matrices \cite{Jolliffe2002}. This technique has been successfully applied in many diverse areas, ranging from finance \cite{Plerou2002} to molecular simulation \cite{Stein2006}. PCA accomplishes the task of dimensionality reduction by taking a subset of the orthogonal basis for the correlation matrix $\mathbf{E}$ \cite{Jolliffe2002}. The first principal component corresponds to the eigenvector with the highest eigenvalue, providing the direction where the data is maximally spread out i.e. explains the most variance of the system. Each subsequent principal component has a lower eigenvalue and thus explains a lower fraction of the total variation of the system. Therefore, we can reduce the dimensionality of the correlation matrix by taking a subset of principal components, hoping to encode most of the total variance of the data. This subset can be chosen with the help of Random Matrix Theory (RMT) \cite{Bun2017}, which studies the properties of matrices drawn from a probability distribution. In our specific context of forming composite indicators in a data-driven way, one could then use the chosen subset of components as a basis for composite indicators.   

In this section, we apply PCA to the correlation matrix $\bm{E}$ on the dataset of \cref{WDI_Data}, finding the distribution of its eigenvalues, using results from RMT to help interpret it. We then analyse the contribution of each topic defined in \cref{AmalgamatingTopics} to the eigenvectors corresponding to the principal components. %For our problem, it would also provide a natural way of constructing the composite indicators by simply using the orthogonal eigenvectors as weights. 

\subsection{Eigenvalue Spectrum} \label{EigenvalueSpectrum}
As is customary in Random Matrix Theory, we fitted the Mar\v{c}enko-Pastur (MP) distribution \cite{Marchenko1967} to the eigenvalue distribution of $\bm{E}$ to discern what part of the eigenvalue spectrum is less likely to be a product of finite-sampling noise. We found that MP does not fit our eigenvalue distribution well, which suggests that there is structure in the whole distribution, as opposed to just its right tail. We shuffled the data to destroy all correlations between indicators, and obtained an eigenvalue distribution that fitted the MP near perfectly. These findings suggest that choosing only a subset of the principal components obtained by PCA is likely to discard relevant information. In other words, this is a clue that PCA might be unsuitable to reduce dimensionality on this dataset. For a more detailed discussion of the procedures in this subsection, we refer to the Supplementary Information \cref{EigenvalueSpectrumSupplementary}.

\subsection{Eigenvector Interpretation} \label{EigenvectorInterpretation}
We investigate what the interpretation of the eigenvectors is by calculating the contribution of each of the $G$ topics from \cref{AmalgamatingTopics} that divide the indicators. This will reveal the structure with respect to topics of the principal components so we can see if they are dominated by one specific topic. The analysis will also be particularly relevant for the earlier principal components that are the main contributors to the variance of the system, which will bring to the surface any topics which are more significantly contributing to development.

\begin{figure}
\centering
\includegraphics[width=0.7\textwidth]{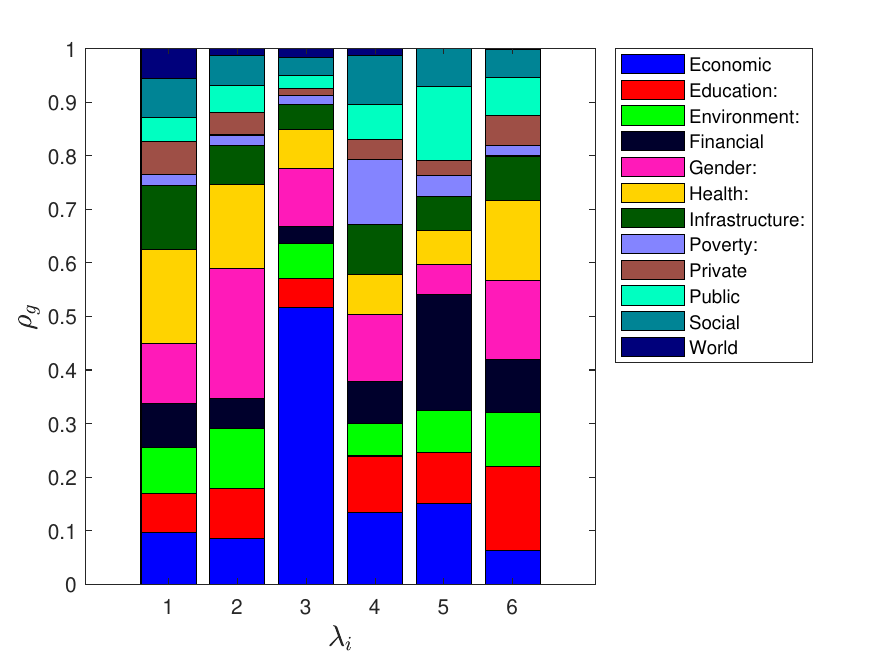}
\caption{Bar chart of the $\rho_{g}$ defined in \cref{RhoG} for the top $6$ principal components of $\bm{E}$ using the $12$ topics of the indicators in \cref{AmalgamatingTopics}. The legend corresponds to these $12$ topics.}
\label{fig:RhoG_6PCs}
\end{figure}

Specifically, we project the eigenvectors $\bm{v}_{i}$ of $\bm{E}$ onto the $G$ topics which divide the indicators that we defined in \cref{WDI_Data} using the projection matrix $\bm{P}$ with entries
\[P_{ig} = \begin{cases}
						1/N_{g} & \text{if $i$ is in topic $g$} \\
						0       & \text{else}\ ,	
						\end{cases}
\]
where $N_{g}$ is the number of indicators that are part of topic $g$. From this, for every we can define $\bm{\rho}_{i}$, which is $G$-dim vector with entries $\rho_{g,i}$, and is computed as 
\begin{equation}
\bm{\rho}_{i}=\gamma_{i}\mathbf{P}\mathbf{v}_{i} \ , \label{RhoG}
\end{equation}
where $\gamma_{i}$ is the normalisation constant $\sum_{g=1}^{12}\rho_{g,i}$. Each entry of $\bm{\rho}_{i}$ gives the contribution of the $g$-th topic to the $i$-th eigenvector. As an example, we plot $\bm{\rho}_{i}$ for the top $6$ principal components in \cref{fig:RhoG_6PCs}.

In \cref{tab:RhoG_PVal}, we report the one-sided p values of $\rho_{g}$ for testing against the null hypothesis that the contribution from the topic to the principal component is random using the procedure detailed in Supplementary Information \cref{RhoG_StatTest}. The bolded values are those below the $5\%$ significance level where we reject the null hypothesis. By looking at \cref{fig:RhoG_6PCs} and \cref{tab:RhoG_PVal}, we see that for the first principal component although other topics contribute to the largest eigenvalue, the statistically significant contributions come from the Health and Infrastructure. Similarly for the second principal component the Environment, Health and Gender topics make a statistically significant contribution, and for the third principal component only the Economic related indicators make a significant contribution.

\begin{table}
  \centering
	\caption{One sided p values of the $\rho_{g}$ defined in \cref{RhoG} using the $12$ topics defined in \cref{AmalgamatingTopics}. The values were calculated using the procedure detailed in the Supplementary Material section 3 and the null hypothesis used is that the $\rho_{g}$ is random. The bolded values are the ones which are below the $5\%$ significance level.}
  \adjustbox{width=0.9\columnwidth,center}{
    \begin{tabular}{|c|c|c|c|c|c|c|}
    \hline
          & \multicolumn{6}{c|}{pval} \\
    \hline
    'Economic' & 0.109263 & 0.503665 & \textbf{0} & \textbf{4.60E-07} & \textbf{0} & 0.965214 \\
    \hline
    'Education' & 0.818697 & 0.237016 & 0.987429 & 0.05904 & 0.224225 & \textbf{6.90E-07} \\
    \hline
    'Environment' & 0.447607 & \textbf{0.01599} & 0.932927 & 0.959271 & 0.689149 & 0.110319 \\
    \hline
    'Financial' & 0.543933 & 0.958713 & 0.999491 & 0.646416 & \textbf{0} & 0.200756 \\
    \hline
    'Gender' & 0.140525 & \textbf{3.68E-06} & 0.175278 & 0.07279 & 0.849132 & 0.016654 \\
    \hline
    'Health' & \textbf{0} & \textbf{0} & 0.838467 & 0.788524 & 0.952838 & \textbf{0} \\
    \hline
    'Infrastructure' & \textbf{0.043397} & 0.733465 & 0.983559 & 0.312448 & 0.854746 & 0.517028 \\
    \hline
    'Poverty' & 0.999458 & 0.999486 & 0.999633 & 0.079216 & 0.98248 & 0.999533 \\
    \hline
    'Private' & 0.958778 & 0.99881 & 0.999667 & 0.999483 & 0.999664 & 0.983491 \\
    \hline
    'Public' & 0.996577 & 0.990442 & 0.999663 & 0.896429 & 0.000541 & 0.812393 \\
    \hline
    'Social' & 0.80534 & 0.97684 & 0.999635 & 0.287871 & 0.853167 & 0.988461 \\
    \hline
    'World' & 0.41984 & 0.709084 & 0.656062 & 0.697268 & 0.923661 & 0.909286 \\
    \hline
    \end{tabular}%
    }
  \label{tab:RhoG_PVal}%
\end{table}%

\begin{figure}
\centering
\includegraphics[width=0.7\textwidth]{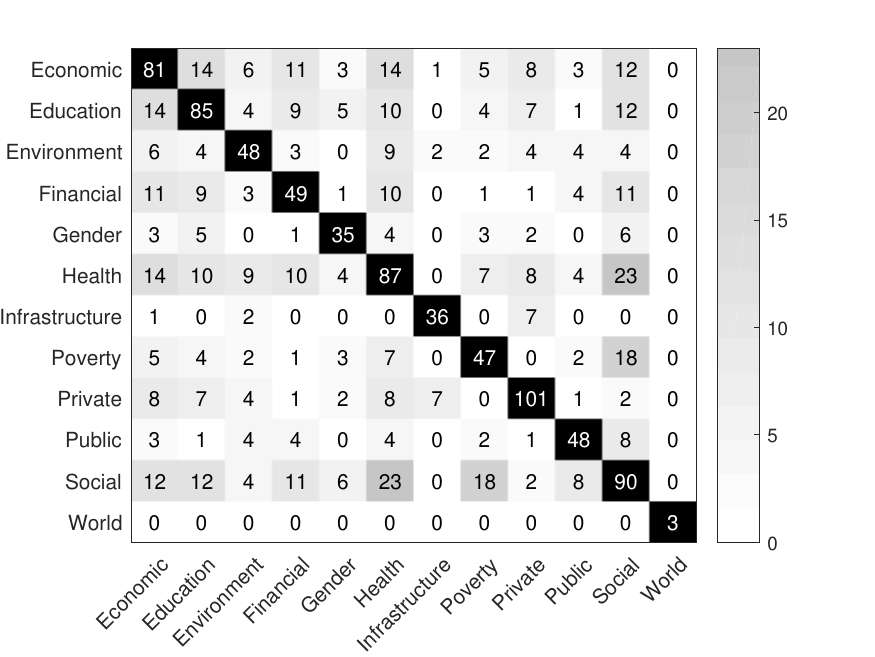}
\caption{Grey scale map with the off diagonal entries giving the total number of times that the topics labelled by the corresponding row and column are simultaneously statistically significant across all principal components. A p-value of $0.05$ with a Bonferroni correction is used so that the new p-value is $0.05/N$ or $3.45\times 10^{-3}$. The diagonal entries are the number of times a single topic is significant at the same p-value across all principal components.}
\label{fig:Graymap_Pval_Double}
\end{figure}

We have also plotted the number of times two topics are simultaneously significant across all principal components in \cref{fig:Graymap_Pval_Double}, with a darker grey indicating a higher number of times this occurs. We use a $5\%$ p value with a Bonferroni correction of $N$, giving the actual p value used to be $3.45\times 10^{-3}$. The black diagonal terms give the number of times a single topic is significant across all principal components using the same p value. If the indicators could be neatly divided into topics then we should see no interaction between them so that \cref{fig:Graymap_Pval_Double} will look almost like a diagonal matrix. In fact, we see that some of the off diagonal elements are quite large relative to the diagonal elements e.g. Education vs Economic and Social vs Health, which indicates that the topics are indeed interacting. We can therefore conclude from this analysis that there is not a clear, single topic that contributes more than others and that in fact statistically significant contributions can come from different topics that combine in different ways.

%It is clearer now that these different topics can interact in numerous different ways.      
%This can be rationalised since a healthy population requires a high level of infrastructure, and likewise better infrastructure will increase the health of a population \cite{Agenor2008}. 

This has implications for composite indicators aiming to capture some particular aspect of development such as GCI and HDI since it suggests that the inclusion of certain indicators which focus on other aspects of development might improve the quality of the composite indicator. Conversely, some indicators may actually not be representative of the aim of the composite one, which means including it would add no information with respect to the aim of the composite indicator whilst also simultaneously increasing complexity. These problems have also been mentioned in \cite{Mazziotta2019} and \cite{Mishra2008}, where it has been shown that potentially important indicators are ignored when forming composite indicators from PCA. Overall, we can conclude that whilst the principal components indicate that the correlations between indicators contains interesting structure, it is difficult to use PCA to form new composite indicators. This means we must turn to other methods to achieve both of these goals.       

%\begin{figure}
%\centering
%\includegraphics[width=0.7\textwidth]{Figures/RhoG_6PCs.eps}
%\caption{Bar chart of the $\rho_{g}$ defined in \cref{RhoG} for the top $6$ principal components of $\bm{E}$ using the $12$ topics of the indicators in \cref{AmalgamatingTopics}. The legend corresponds to these $12$ topics.}
%\label{fig:RhoG_6PCs}
%\end{figure}

%\begin{figure}
%\centering
%\includegraphics[width=0.7\textwidth]{Figures/Graymap_Pval_Double.eps}
%\caption{Grey scale map with the off diagonal entries giving the total number of times that the topics labelled by the corresponding row and column are simulateneously statistically significant across all principal components. A p-value of $0.05$ with a Bonferroni correction is used so that the new p-value is $0.05/N$ or $3.45\times 10^{-3}$. The diagonal entries are the number of times a single topic is significant at the same p-value across all principal components.}
%\label{fig:Graymap_Pval_Double}
%\end{figure}

%However, we do not rely on PCA in the formation of the CDCIs. Moreover, we do include the information coming from lowly inter-correlated indicators since every indicator is a member of a cluster, which forms an individual CDCI. This information and clarification has now been added on page $5$ and $8$.

\section{Interpretation of the clustering from the DBHT} \label{DBHT_interpret}
This section analyses the relationships between indicators in a data driven way where we make as little assumptions about the structure of the data as possible. In this way, we can develop an interpretation and partition of the indicators which is consistent with the data. In the previous section, we showed that this is not possible with PCA and by dividing indicators based on their a priori topic given in \cref{AmalgamatingTopics}.

Hierarchical clustering algorithms \cite{Bishop2006}, which group together data with similar properties in a hierarchical fashion, and their associated network filtering techniques will help in this respect. This is because we can consider information from all indicators. Hierarchical clustering is advantage in our context since it has been shown with many other different kinds of data originating from complex systems that data is organised hierarchically \cite{Ravasz2003,Corominas2013}. Therefore it seems natural to apply a hierarchical clustering algorithm to discover this structure. Once we apply the clustering algorithm on the correlation matrix, we have a natural way of accomplishing dimensionality reduction by using one variable to describe each cluster of nodes, with the collection of clusters forming the reduced correlation matrix. The ones associated to network filtering algorithms leverage on the topological properties of the filtered network.

We shall use the PMFG network filtering technique because it is able to retain a higher amount of information about the system than the MST. This is because it preserves a greater number of links of the original network and in fact contains the MST as a subgraph \cite{Tumminello2005}. This is important for us since the MST is a tree and thus contains no loops, whereas the PMFG contains $3$ and $4$ cliques, and we would like to avoid discarding relevant information about the relationship between indicators. For the PMFG, the associated clustering algorithm is the DBHT algorithm, which takes advantage of the $3$ clique structure of the PMFG. There has been objections to the use of cluster analysis due mainly to the distance between points as a metric \cite{Jain1999,Wang2002} since it does not measure the similarity between variables. Instead, the DBHT algorithm forms the clustering by measuring forming a distance matrix directly from the similarity matrix, which encodes the relationships between variables. In our case the similarity matrix is $\bm{E}$, with the entries of the distance matrix $D_{ij}$ commonly defined as \cite{Mantegna1999,Mantegna2000,Musmeci2015}  
\begin{equation}
    D_{ij}=\sqrt{(2\left(1-E_{ij}\right))} \ .
\end{equation}
$D_{ij}$ therefore measures how far two indicators are in terms of their correlation.    

The main advantage of using the DBHT algorithm is that it does not need prior input into the number of clusters, making it preferable over other clustering algorithms so that we can make a-posteriori comparison with less assumptions \cite{Song2012,Musmeci2015}. This is important for us since we want to uncover the structure of correlations between indicators, making as few assumptions as possible on this same structure. Furthermore, the DBHT algorithm has been shown to retrieve a higher amount of meaningful information\cite{Musmeci2015}.
%In this section, we make a more detailed analysis of whether the indicators can be divided into their topics by applying the DBHT algorithm to $\bm{E}$ in \cref{DBHT_results}. 

In this section, we investigate whether the indicators can be divided into their topics by applying the DBHT algorithm to $\bm{E}$ in \cref{DBHT_results}. Then by analysing clusters individually, we look for their dominating topics and what their possible interpretation is in \cref{DBHT_similarity}.   

\subsection{DBHT results and interpretation} \label{DBHT_results}

\begin{figure}
\centering
\includegraphics[width=0.7\textwidth]{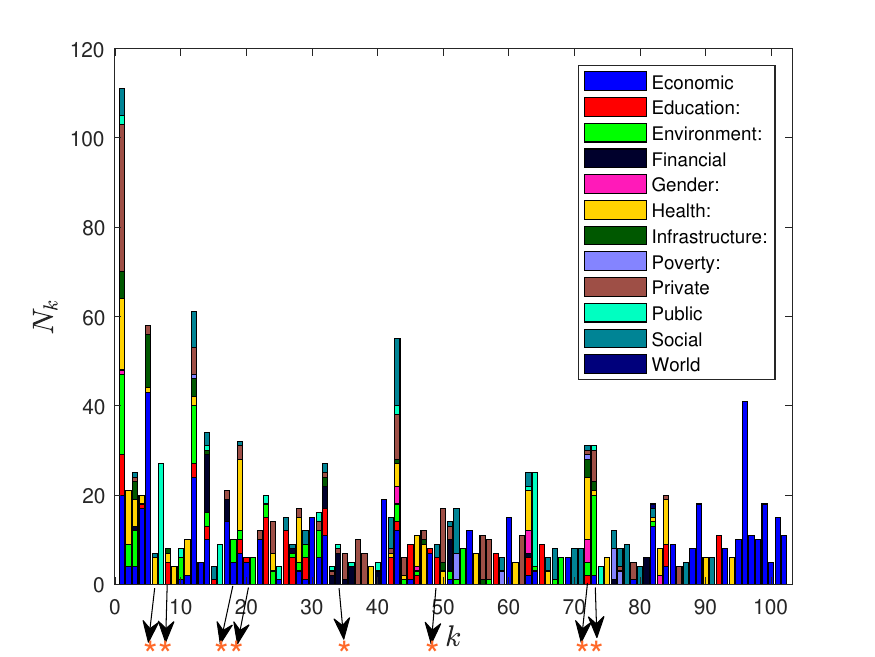}
\caption{The cluster label $k$ versus the number of indicators in each cluster $N_{k}$. Each bar is divided into the number of indicators in cluster $k$ which belong to each topic, with each colour corresponding to each topic according to the key on the right.The arrows with the orange asterisks point to clusters (starting from the left) $6, 8, 18, 20, 34, 49, 72, 73$, which are used in \cref{fig:panel_timelapse} and \cref{fig:panel_2}.}
\label{fig:Cluster_Topic_Dist}
\end{figure}

We apply DBHT to $\bm{E}$. It identifies a total of $K=102$ clusters which we label $k=1,...,K$, significantly more than the $G$ preassigned topics, with an average cluster size of $14.2$. In \cref{fig:Cluster_Topic_Dist} we summarise the clustering labels obtained from DBHT and its topic composition, with the height of the bar representing the number of indicators in each cluster $N_{k}$. Each bar is further divided by colours which represent how many indicators belong to that particular topic. \Cref{fig:Cluster_Topic_Dist} shows that cluster sizes are highly heterogeneous - the biggest cluster has $111$ indicators versus the smallest with $4$.

We can see that some clusters are dominated by certain topics - for example cluster $41$ is dominated by economic indicators and in particular indicators related to countries' current account balance and external balance of trade.
At the same time however, there are also some clusters which are instead a mixture of topics but still have an interpretation based on the indicators contained within them such as cluster $72$, which contains indicator from disparate topics. A closer inspection reveals that this cluster is made of  indicators about access to electricity, railways size, primary and secondary education expenditure, health-related indicator such as HIV incidence and hepatitis immunization, access to sanitation facilities, prevalence of underweight children, number of women who justify a husband's beatings, and the Gini index. All these measurements can be easily used to characterize underdeveloped countries \cite{Winkler2011,Garcia2006,Smith2000,Ravallion1997,Bose2007,Gupta2008,Montgomery2007}%\todo{it's common knowledge, references needed tho}.

Another interesting fact that we can observe from the data is that the cluster $5$ contains very important economic indicators such as GDP per capita, value added contributions of agriculture, industry, manufacturing, services and trade, and also imports/exports of goods and services as a fraction of GDP. This same cluster also contains indicators directly related to measuring the innovation output of a country such as patent, trademark and industrial design applications, suggesting that innovation is an important factor in economic development. We can interpret this by realising that innovation led growth increases productivity through the accumulation of knowledge obtained via education, new products or better processes \cite{Romer1990,Aghion2010}.

Many other interesting clusters are found, such as number 11, which seems to relate to underdevelopement with it contents relating to life expectancy, foreign aid, drinking water availability, fertility rate, and percentage of women married before the age of 18. Cluster 21 puts together CO2 emissions, alternative and nuclear energy, combustible renewables and waste, hydroelectric sources prevalence and power distribution losses. Cluster 101 describes the distress status of a country's debt, including indecators about how much of it has been rescheduled or forgiven.

%\begin{figure}
%\centering
%\includegraphics[width=0.7\textwidth]{Figures/Clusters_Topic_Dist_PMFG.eps}
%\caption{The cluster label $k$ versus the number of indicators in each cluster $N_{k}$. Each bar is divided into the number of indicators in cluster $k$ which belong to each topic, with each colour corresponding to each topic according to the key on the right.}
%\label{fig:Cluster_Topic_Dist}
%\end{figure}

\subsection{Similarity of the DBHT clustering with the topics} \label{DBHT_similarity}
Once we have established that each of the clusters has an economic meaning, we quantify how much the clustering outputted by the DBHT in \cref{fig:Cluster_Topic_Dist} is similar to the clustering based on topics. Therefore, we will be able to see overall how close the two divisions of topics are in a quantitative way. We do so using the Adjusted Rand Index (ARI), which is $1$ if there is a perfect agreement, $-1$ if there is an anti-agreement and $0$ if there is no agreement \cite{Rand1971}. It has been successfully used in \cite{Musmeci2015}. Computing the ARI to compare the output of the DBHT algorithm and the topics distribution, we find that this value is $0.0456$ i.e. quite close to $0$, which corroborates our previous conclusion that overall the clustering of the data is not in general linked to that based on topics.

The analysis can also be made at local level by seeing if any topics have a significant presence in each cluster. In this way, it allows us to also to see locally if more than one topic might be present in each cluster, which is important since whilst on a global level there may not be much similarity . Practically, this is achieved by using the procedure proposed in \cite{Tumminello2011}. Specifically, we test statistically, using a one sided test, the null hypothesis that a cluster $k$ from the DBHT and the $g$-th topic have $m$ common elements is random. Under the null hypothesis, this distribution is hyper geometric. If the null hypothesis is rejected, it means that statistically we say that the $g$-th topic is \emph{overexpressed} in cluster $k$. We apply this procedure to each of the DBHT clusters and topics using the p value of $8.17\times 10^{-6}$ (which is $0.01$ with a Bonferoni correction \cite{Feller2008} of $1/2 KG$), recording the number of overexpressed topics in each cluster. The results of this procedure reveals that whilst a majority of clusters have one or two topics overexpressed, there are a total of $49$ clusters which have no overexpressed topics. These particular clusters of indicators still have an economic meaning. For example, cluster $32$ contains indicators relating to tertiary education such as pupil to teacher ratio in tertiary education and completed education at a tertiary level, which belong the education topic. However, it also contains indicators such as scientific and technical journal articles, which is classed as relating to infrastructure. These indicators may be linked e.g. because scientific articles are usually always published by authors with at least a tertiary level education. This confirms our conclusions that overall at a system wide and local level, the clustering of the data does not reflect the information given by the topics, suggesting that indicators do not necessarily correlate with other indicators of the same type.

\section{Deriving new composite development indicators from DBHT} \label{CompInds_DBHT}
In the previous section we showed that the distribution of topics amongst the indicators is not an accurate description of the data and may miss key information about the relationship between different classes of indicators. This means that composite indicators based on this premise such as the HDI or the WEF-GCI infrastructure pillar may not be the best way of combining indicators. We want to propose a new set of data driven composite development indicators which can encapsulate this new information based on the results given in \cref{DBHT_results}. In doing so, we would overcome traditional problems faced when forming composite development indicators, mainly on how and which indicators we should aggregate. This section is dedicated to describing a way of using the results in section \ref{DBHT_interpret} to derive a novel set of cluster driven composite development indicators.

To define each composite indicator we shall use the set of clusters from DBHT given in \cref{DBHT_results}. It provides a natural way to select the indicators to combine for our composite indicators since each cluster contains indicators which share similar properties, and also has an economic interpretation as highlighted in section \ref{DBHT_interpret}. Hence, aggregating information for indicators which are members of the same cluster enables us to simply and efficiently summarise the economic information contained within them. In contrast with PCA \cite{Mazziotta2019,Mishra2008}, since we rely on the DBHT clustering, we automatically include information coming from all the indicators since every indicator is a member of some cluster that defines its respective composite version. Condensing the complimentary insight offered by indicators in the same cluster also overcomes the need to make 'educated' assumptions about which indicators are to be combined that other alternative composite indicators often use \cite{Sagar1998}. In this way, we can more clearly see the overall behaviour of each set of indicators in cluster $k$ by using the corresponding composite indicator value as a proxy. DBHT also is significantly advantageous in this respect since it requires no prior input of the number clusters (and thus number of composite indicators) needed to describe the properties of the data) \cite{Song2012,Musmeci2015}. Potentially, opposite polarities of indicators in the same cluster could make the interpretation of the value of any composite indicator formed from the same cluster problematic \cite{Mazziotta2019}. However, an advantage of forming composite indicators from the DBHT is that the clusters from the sparse correlation network PMFG, which we calculated has only $1\%$ of its non negative entries as negative. Hence, the indicators in the same cluster have same polarity.

\subsection{Method used to calculate the composite indicators}
%We must also establish a procedure on how we should aggregate the indicators that we are combining. This is important since there may be some indicators within each cluster which are more crucial to determining the behaviour of other indicators through their greater influence.
Here, we shall define the method used to calculate the new composite development indicators based on the results of \cref{DBHT_results}. In the $k$ composite indicator we want to capture the average behaviour of all indicators in that cluster. Therefore, we aggregate the indicators in cluster $k$ by using the median value across all indicators within this cluster. We can do this because the indicators are standardised and normalised via the procedure in the Supplementary Information \cref{Distribution_regularisation}. This forms composite indicator $k$, $I_{k}$, defined as 
\begin{equation}
I_{k}=\mathrm{median}_{i \in \text{cluster } k}\mathbf{X} \ ,
\end{equation}
where the notation $i \in \text{cluster } k$ indicates that we only take indicators $i$ that are members of cluster $k$. An advantage of using the median over the arithmetic mean or even a weighted mean is that the median is more robust to outliers. The median is a valid measure across the different indicators because the entries of $\bm{X}$ are also standardised, meaning that their scales are all the same. We highlight that we have chosen to use the median for every $k$ since this provides us with a consistent methodology so that the precise details of how $I_{k}$ is calculated do not change for every $k$. This improves some existing methods used in the literature where for example the same indicator may be calculated in different ways for different regions \cite{Huggins2003} meaning that we can make valid comparisons between indicators. Furthermore, a PCA based approach would require the use of weights, which has been argued to be hard to interpret economically \cite{Mazziotta2019, Somarriba2009}. Instead, the median used does not require weights as an input to form the composite indicators we propose. We use this method to calculate the set of $I_{k}$, giving $102$ indicators in total and call this set of cluster driven composite indicators (CDCIs). %As an example we show this for cluster [cluster], for [country]. We see an increasing trend , which matches the increase in development that has been has been observed previously.    

%For this, we shall use the submatrix of $\mathbf{E}$ which only refers to members of that cluster, $\mathbf{E}^{(k)}$ computed as 
%\begin{equation}
%\mathbf{E}^{(k)}={E}_{(i,j) \in \text{cluster } k} \ ,
%\end{equation}
%where the notation ${...}_{(i,j) \in \text{cluster } k}$ refers to keeping only rows and columns of the original $\mathbf{E}$ which correspond to cluster $k$. This means that $\mathbf{E}^{(k)}$ only contains the structure of the intra-correlations of a cluster. $\mathbf{E}^{(k)}$ is therefore a good candidate to study to find $\xi_{i}$ since it must normalised across only those indicators which part of that cluster $k$. If we calculate the eigenvalues of $\mathbf{E}^{(k)}$, then the first principal component will give the highest contribution to the variance to restricted system consisting of only indicators that are in the cluster. The corresponding normalised eigenvector $\mathbf{v}^{(k)}$ is a natural candidate for $\xi_{i}$ as it gives the contribution of each of these indicators to the first principal component, hence we set $\xi_{i}=\mathbf{v}^{(k)}$.

\section{Using the CDCIs to understand country development} \label{CountryDevelopment}
One of the main uses of indicators is to track the country development of a country to assess its progress. This is important since it gives an idea of what has been achieved in terms of country development and where to focus policy changes to affect country development positively. One can also use indicators to compare countries either pair wise or globally. On the last point, CDCIs can be useful because the methodology used to compute them is not reliant on subjective, country dependent criteria, which means we can make fair comparisons between different values of the CDCIs for different countries at different times. By comparing the CDCIs with each other for all countries, we can therefore investigate whether they can be used to assess the development of a country. The main CDCIs we shall concentrate on are $6, 8, 18, 20, 34, 49, 82, 73$, which are marked by orange asterisks in \cref{fig:Cluster_Topic_Dist}.

\begin{figure}
    \centering
    \includegraphics[width=0.9\textwidth]{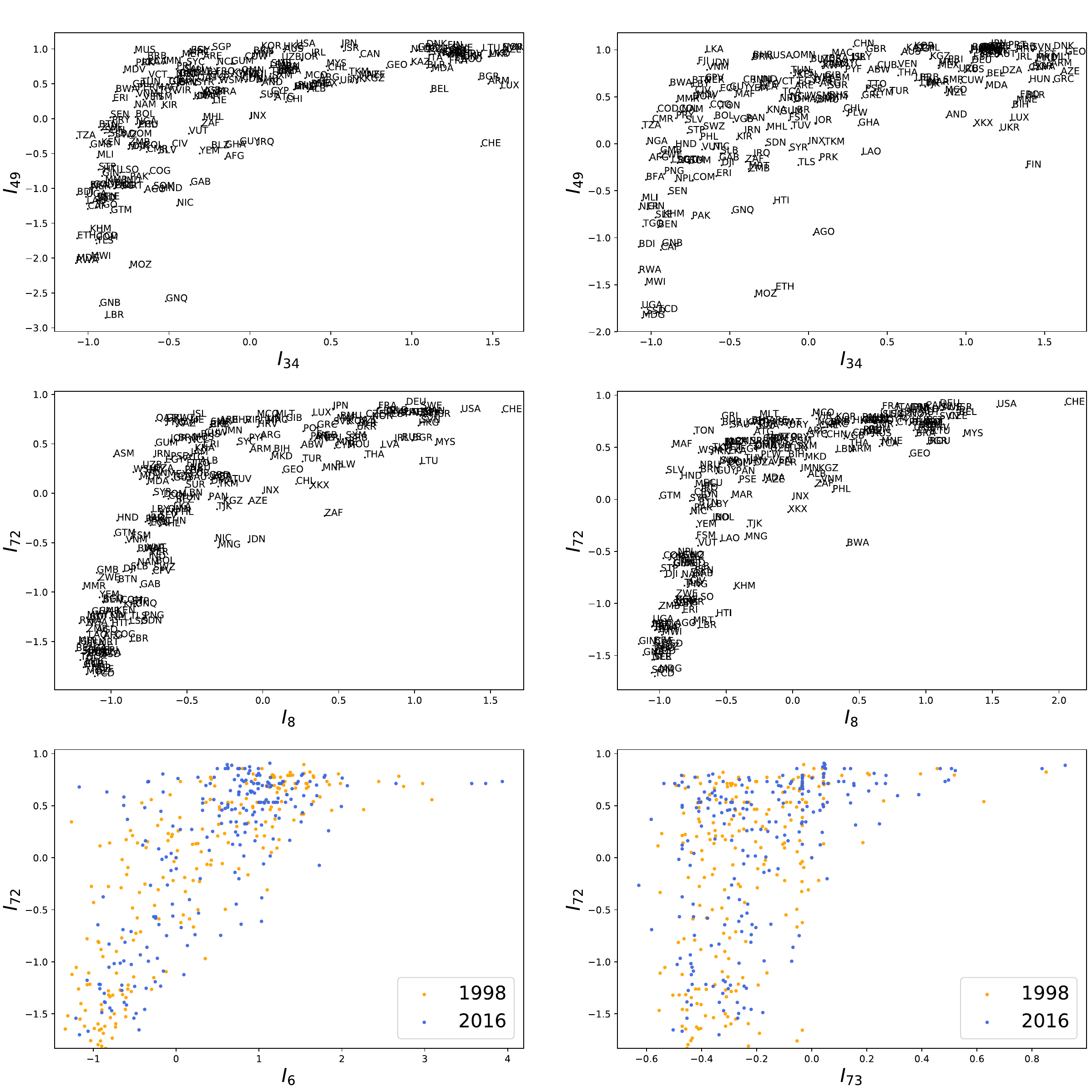}
    \caption{(Top panel) $I_{34}$ vs $I_{49}$ for $1998$ on the left and $2016$ on the right. (Middle panel) The same but with $I_{8}$ vs $I_{72}$. (Bottom panel) On the left we have $I_{6}$ vs $I_{72}$ for $1998$ in orange and $2016$ in blue. On the right is the same but for $I_{73}$ vs $I_{72}$.}
    \label{fig:panel_timelapse}
\end{figure}

In \cref{fig:panel_timelapse}, we provide some examples of comparisons between different indicators. Overall, we remark that all the plots have a 'hockey-stick' shape. We can specifically see for the plots in the top two panels the vertical leg of the hockey-stick shape is made of developing countries, whilst the horizontal leg, indicating a saturation effect, is made of developed countries. This is interesting since it suggests a country level transition from a group consisting of developing nations to one with developed nations. In fact, this further supports the so called two regime hypothesis \cite{Pugliese2017,Cristelli2018}, where countries below a barrier struggle to develop consistently, which corresponds to nations in the vertical leg of hockey-stick shape. Countries that overcome this barrier have or are experiencing high growth in development, which are represented by the horizontal leg of the hockey-stick shape. This transition can clearly be observed to be consistent across time from the bottom panel comparing $I_{6}$ and $I_{72}$ and $I_{73}$ and $I_{72}$, with the years $1998$ and $2016$ overlayed.

As a consequence of the consistency of the observed hockey-stick shape, we can make interesting observations regarding the particular pair of CDCIs being plotted. Specifically in the top panel of \cref{fig:panel_timelapse}, we plot $I_{34}$ against $I_{49}$, where the former represents those who have use mobile and banking services and the latter come from primary school statistics, a key signature of development \cite{Keller2006}. We see here that it seems that the vertical leg access to mobile phone technology can, for developing countries, characterise their development. Past a certain point however, the concavity changes, suggesting that access to mobile phones becomes less able to distinguish between countries' development. In this region, we have already remarked that they are mostly developed nations, who have a saturation in mobile phone access due to their higher average income. Likewise, we see in the middle panel, which corresponds to $I_{8}$ (secondary school enrollment) and $I_{72}$ (recalling from \cref{DBHT_results} that this represents underdevelopment), that secondary school enrollment can initially also be used to characterise development. However after a certain level of development, secondary school enrollment saturates in these developed countries, meaning it can no longer be used in this way.

\begin{figure}
    \centering
    \includegraphics[width=0.9\textwidth]{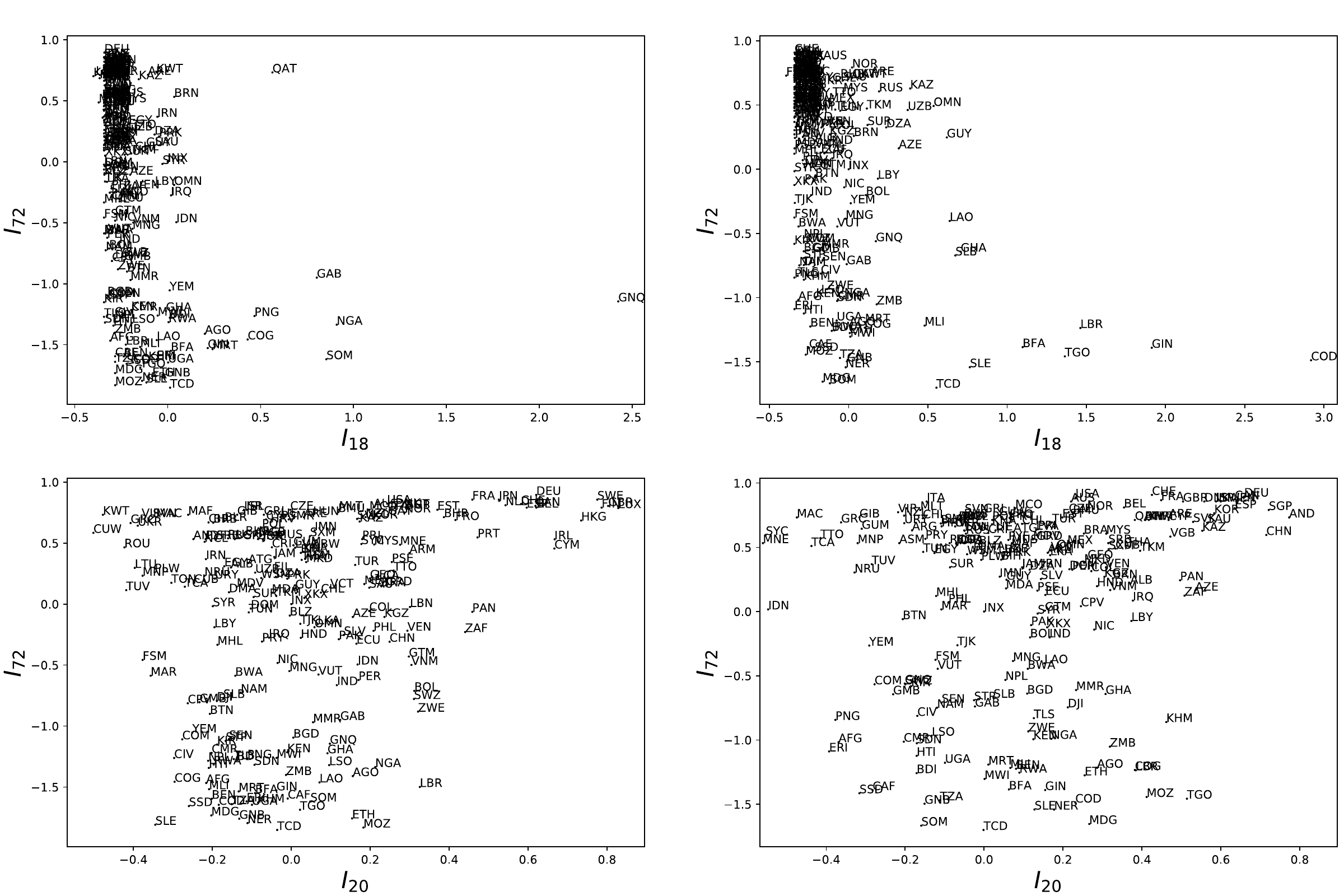}
    \caption{(Top panel) $I_{18}$ vs $I_{72}$ for $1998$ on the left and $2016$ on the right. (Bottom panel) The same, but for $I_{20}$ vs $I_{72}$.}
    \label{fig:panel_2}
\end{figure}

However, not all relationships between certain CDCIs are hockey-stick shaped. Indeed, we can see this from \cref{fig:panel_2} which plots $I_{18}$ vs $I_{72}$ for $1998$ on the left and $2016$ on the right in the top panel and the same but for $I_{20}$ vs $I_{72}$ in the bottom panel. For the top panel, $I_{18}$ is a CDCI that represents natural resource abundance, whilst again we recall from \cref{DBHT_results} that $I_{72}$ corresponds to underdevelopment. We notice from the plots in the top panel that most of the countries with higher abundance of natural resources are underdeveloped countries. This reminds of the so called 'resource curse' \cite{Ross1999}, where resource-rich nations with inefficient governments are often underdeveloped.

Additionally in the bottom panel of \cref{fig:panel_2}, we plot $I_{20}$ vs $I_{72}$. $I_{20}$ corresponds to the amount of flow of foreign direct investment (FDI). Interestingly, we can observe an intriguing relationship between underdevelopment and FDI in the plots. All underdeveloped nations tend to have low FDI, which can be interpreted as being perceived with a low investment potential by foreign investors. However, countries which are not underdeveloped may have both high or low FDI, for example Poland, Kuwait and Uzbekistan are all more highly developed countries that have a low FDI. Attractiveness to foreign investments is not directly correlated to a country's level of developement.

%\begin{figure}
%    \centering
%    \includegraphics[width=\textwidth]{Figures/panel_timelapse.pdf}
%    \caption{Comparison between different CDCIs. (Top panel) $I_{34}$ vs $I_{49}$ for $1998$ on the left and $2016$ on the right. (Middle panel) The same but with $I_{8}$ vs $I_{72}$. (Bottom panel) On the left we have $I_{6}$ vs $I_{72}$ for $1998$ in orange and $2016$ in blue. On the right is the same but for $I_{73}$ vs $I_{72}$.}
%    \label{fig:panel_timelapse}
%\end{figure}

%\begin{figure}
%    \centering
%    \includegraphics[width=\textwidth]{Figures/panel_2.pdf}
%    \caption{(Top panel) $I_{18}$ vs $I_{72}$ for $1998$ on the left and $2016$ on the right. (Bottom panel) The same, but for $I_{20}$ vs $I_{72}$.}
%    \label{fig:panel_2}
%\end{figure}

\section{Deriving influential indicators by using PMFG} \label{Influential_PMFG}
%Another class of dimensionality reduction techniques rely on network filtering. These procedures view $\mathbf{E}$ as a network, with each node representing an indicator in our case and each edge representing a non-zero correlation between two indicators. Information filtering techniques aim to take the dense correlation network $\mathbf{E}$ and reduce it to a sparse version by extracting the most important interactions. They do this by maximising the correlation subject to a topological constraint on the network. The induced sparsity of the network will make important hidden structures more visible, enabling us to assess and interpret the interactions between indicators more easily. One successful example of such a technique is the Minimum Spanning Tree (MST), which maximises correlations subject to the constraint that the resulting network is a fully connected tree. It has been successfully applied to problems in finance \cite{Mantegna1999} and biology \cite{Trapnell2014}.
One can imagine that there could be nodes that are very important to the structure of the correlation network than others, implying that these same nodes could be highly influential in the analysis of the development of countries. This would be very interesting for our purposes since they could provide a direct way to form a reliable reduced set of development indicators that would automatically overcome any problems associated with calculating composite versions. More specifically, one could take a subset of the top most influential indicators since these indicators' influence have an aggregate, over-arching influence on all other indicators, and thus the structure of interactions in the system. 

In this section we shall apply the PMFG to $\bm{E}$ and identify system wide important indicators. For this purpose, information filtering is useful because it neatly transfers the problem of identifying influential indicators as finding a ranking of important nodes in the network, for which there exist several so called network centrality measures. We choose to use PageRank \cite{Page1999}, which has proven successful in ranking scientists and webpages \cite{Page1999,Liu2005}, to identify the most system wide influential indicators that affect the network. In PageRank, we rank nodes of networks on their importance based on the probability of a random walker landing on a particular node \cite{Page1999} with higher values indicating that the node has more importance.

We find the PMFG of $\bm{E}$. The output network of this visualisation can be seen in the Supplementary Information \cref{PMFG_network}. Then we apply PageRank to the PMFG of $\bm{E}$, displaying an example of the top $9$ indicators in \cref{tab:PageRank}.

\subsection{Interpretation of the PageRank identified indicators} \label{PMFGAnalysis}
From \cref{tab:PageRank} we observe that there are some indicators which we would expect to be in this ranking: for example GDP measures the value of goods and services an economy produces, and is widely used as a primary development indicator \cite{Lepenies2016}. Central government debt has also been linked to economic development since high levels of debt can drag growth rates down \cite{Checherita2012}. However, it is interesting to see that mobile cellular subscriptions to be the top ranking indicator, especially considering our comments in \cref{CountryDevelopment} that mobile banking can be used to track the development of countries. This is an interesting result since there are many papers in computational socioeconomics which use mobile data as metric of an average citizen's socioeconomic status due to vast information it can encode \cite{Blumenstock2010,Blumenstock2012,Mehrotra2012,Gutierrez2013,Gao2019}. In fact, it has been shown for example that mobile data is correlated with household expenditure \cite{Blumenstock2010} and poverty \cite{Smith2013}, and reveal gender inequality \cite{Mehrotra2012}. This may be because having a mobile cellular subscription requires a number of milestones in the development of a country e.g. a healthy enough population to make use of them, the education to know how to use them, the relevant infrastructure such as phone masts that can reach all parts of the population.

We have also investigated whether particular topics are overexpressed within the top $102$ (chosen because this is the number of clusters identified by the DBHT) PageRank indicators by applying the same hypothesis test used in \cref{DBHT_similarity}. We find that no topic is overexpressed within this subset of indicators, which again corroborates our conclusion that no single topic is more influential than the other.

\begin{table}
  \centering
	\caption{The names of the top $9$ influential indicators based on PageRank in the second column and their actual PageRank values in the first column.}
	\begin{adjustbox}{width=0.9\columnwidth,center}
    \begin{tabular}{|c|c|}
    \hline
    \multicolumn{1}{|l|}{pagerank} & names \\
    \hline
    0.006764 & Mobile cellular subscriptions \\
    \hline
    0.004666 & Share of tariff lines with specific rates, manufactured products (\%) \\
    \hline
    0.003717 & Children in employment, wage workers, male (\% of male children in employment, ages 7-14) \\
    \hline
    0.003706 & Unemployment, male (\% of male labor force) (national estimate) \\
    \hline
    0.003129 & Mobile cellular subscriptions (per 100 people) \\
    \hline
    0.002939 & Central government debt, total (\% of GDP) \\
    \hline
    0.00276 & Share of youth not in education, employment or training, female (\% of female youth population) \\
    \hline
    0.002686 & Population ages 30-34, female (\% of female population) \\
    \hline
    0.002553 & GDP (current US\$) \\
    \hline
    \end{tabular}%
		\end{adjustbox}
	\label{tab:PageRank}
\end{table}%

\section{Performance comparison} \label{PerformanceComparison}
If we reduce all of the indicators in the dataset to the composite ones, we have boiled down the structure of the correlations between indicators to more essential constituents. Therefore, when the set of composite indicators are taken together, they should still be a faithful representation of the original $\bm{E}$ since they are main driving factors behind the structure of correlations. We can use this principle to evaluate the performance of the CDCIs against any alternatives. This section is dedicated to comparing the performance of the CDCIs derived in \cref{CompInds_DBHT} against some alternatives.

For this purpose we propose, as a first approximation, that each indicator can be written as a linear factor model \cite{Thompson2004} of composite indicators. The general linear model is 
\begin{equation}
\bm{X}_{i}=\sum_{k=1}^{K}\beta_{ik}\tilde{I}_{k}+\epsilon_{i} \ , \label{Indicator_FactorModel}
\end{equation}
where $\bm{X}_{i}$ is the $i$th indicator i.e. the $i$th column $\bm{X}$. $\tilde{I}_{k}$ is the $k$-th composite indicator of either the CDCIs or the other alternative schemes of composite indicators. $\beta_{ik}$ is the loading of $i$ for indicator $k$, which measures the sensitivity of $\bm{X}_{i}$ to changes in $\tilde{I}_{k}$. Finally, $\epsilon_{i}$ are white noise terms. \cref{Indicator_FactorModel} is an appropriate approximation to use since firstly, we are using the linear correlation matrix, which means it is intimately related to linear factor models. Note we also have that the number of composite indicators in each of the alternatives used in our comparison must be the same as the number of CDCIs $K$. This is because the size of the indicator set will inevitably affect its ability to describe the correlations, so fair comparison must involve fixing the number of indicators used. 
%Secondly, the first term in \cref{Indicator_FactorModel} can take into account how strongly related indicator $i$ is to its own composite indicator, whilst the second term takes into account any influences that other composite indicators might have. The need for the last term is especially relevant since the composite indicators themselves show significant correlation between each other as we saw in \cref{CountryDevelopment}.
We then use elastic net regression (for details see Supplementary Information \cref{ElasticNet}), which is able to take into consideration the potential correlation between composite indicators, to find $\beta_{ik}$ and $\beta_{ik'}$ for every $i$. 
The performance can then be evaluated on the basis of the error between the linear model and the real indicator values. For this, we define the usual mean squared error of the regression as
\begin{equation}
MSE=\sum_{i=1}^{N}\left(\bm{X}_{i}^{(predict)}-\bm{X}_{i}\right)^{2} \ , \label{MSE_Regression}
\end{equation}
where $\bm{X}_{i}^{(predict)}$ are the predicted values of $\bm{X}_{i}$ using the $\beta_{ik}$ from the elastic regression. The final metric we use the evaluate the performance of the cluster driven composite indicators is
\begin{equation}
ERR=\frac{MSE_{CDCIs}}{MSE_{alt}} \ , \label{ERR}
\end{equation}
where $ERR$ is called the error reduction ratio, $MSE_{CDCIs}$ is the $MSE$ calculated in \cref{MSE_Regression} for the CDCIs. Similarly, $MSE_{alt}$ is the same but for any of the alternative schemes of composite indicators used as a comparison. If $ERR$ is below $1$ (above $1$) then this means that the CDCIs perform better (worse). Also note that of course $ERR$ is bounded below by $0$. 

The choice of alternative schemes of composite indicator, is as follows. We take the top $K$ PageRank indicators that were identified in \cref{PMFGAnalysis}. We choose these particular schemes since they offer the best feasible alternative in forming a basis of composite indicators that have the most influence on correlations system wide. A fourth comparison is also made by randomly selecting $102$ indicators from the columns of $\mathbf{X}$ that provides a benchmark of the performance of the other composite indicator schemes since a set of randomly selected indicators should not be able to reliably incorporate anything from the real correlation network.
%the first $K$ principal components identified through PCA in \cref{PCA} and

We carry out the elastic regression and compute $ERR$ for all alternative schemes of indicators used. For the random benchmark, we repeat and average the results for $100$ different random subsets of indicators. The results are shown in \cref{tab:error_indicators}. We see that in both cases $ERR$ is much less than $1$, indicating that the CDCIs are able to outperform the random benchmark and the PageRank alternative. We can therefore conclude that the CDCIs are more effective at reducing the dimensionality of the dataset the random benchmark and the PageRank alternative.

\begin{table}
  \centering
	\caption{In the first column, the $ERR$ calculated using \cref{ERR} for the benchmark using $102$ random subsets of indicators, which is repeated $100$ times. The second column is the same but instead using $102$ of the most influential indicators, assessed via PageRank.}
    % Table generated by Excel2LaTeX from sheet 'Summary_data'
    \begin{tabular}{|c|c|}
    \hline
    \multicolumn{1}{|l|}{random} & \multicolumn{1}{l|}{PageRank} \\
    \hline
    0.66  & 0.71 \\
    \hline
    \end{tabular}
  \label{tab:error_indicators}
\end{table}                   

\section{Dynamical Analysis} \label{DynamicalAnalysis}
\begin{figure}
\centering
\includegraphics[width=\textwidth]{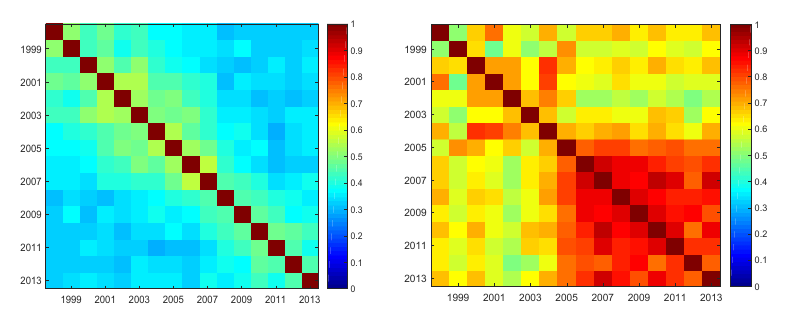}
\caption{(Left)\cref{fig:AR_heatmatrix}a is a heat map of the ARI computed between the DBHT clusterings of indicators for each pair of $16$ rolling time windows of length $4$, with a shift of $1$ year. (Right)\cref{fig:AR_heatmatrix}b is the heat map of the ARI computed between the DBHT clusterings of the CDCIs, using the same parameters for the rolling window as \cref{fig:AR_heatmatrix}a. The colour legend for both plots is on the right.}
\label{fig:AR_heatmatrix}
\end{figure}

Since in the analysis so far we have used the static correlation matrix computed over the whole time period, we should also investigate the dynamic stability of the clusters. We start by splitting the whole time period into $16$ rolling time window of length $4$, with a time shift of $1$ year. For each time window $w=1,...,16$, we calculate the corresponding correlation matrix $\bm{E}^{w}$ and its DBHT clustering. The similarity between each pair of time windows $w$ and $w'$ is then measured by calculating the ARI between their respective DBHT clusterings. The results are shown in the heat map of  \cref{fig:AR_heatmatrix}a. We can see that overall the DBHT clusterings display a high similarity with each other, with a median ARI value of $0.376$, which is high considering that the static clustering does not reflect the topics as argued through the ARI computed in \cref{DBHT_similarity}. This also confirms to some extent our choice for the length of the sliding window since we see that the clusters are stable through time. %\todo{Does this value seem to low to you? Nicolo used a similar procedure and got higher values - see fig. 1/2 of 'Risk diversification: a study of persistence with a filtered correlation-network approach'. Tbh our dataset is way noiser/shorter (each indicator for each country has 19 data points) so probably this comparison isn't fair - just wanted to flag this to you in case. <ORAZIO:> It is low compared to Nicolo's but it's also an unfair comparison. So I think we should mention that the value should be related to the levels of noise and length of our matrix.}

We also used the same procedure and parameters to investigate the dynamic stability of the relationship between CDCIs (except using the correlation matrix and DBHT clustering between the CDCIs). A heat map of the results can be seen in \cref{fig:AR_heatmatrix}b. Again, we see that overall there is a high likeness between the clusterings of the CDCIs in each time window. In fact, the median ARI is even higher at $0.683$. Interestingly, we see that starting from the window covering $2005$ to $2009$, which is the year corresponding to the financial crisis, there is a markedly higher similarity between the clusterings of the CDCIs. This could be explored further. %\todo{Not sure how to interpret this last comment with respect to our results. On the one hand it could indicate that the structure is a new 'phase transition' of more stability? We have to be careful how we use the term phase transition in case it conflicts with our invariant hockey shape stick argument. Maybe you could think of a better interpretation? <ORAZIO:> It definitely marks a shift in the dataset, which could be indicative of some change in the real world. I don't think we should venture into much more interpretation, we should say that in the discussion that this finding will definitely inspire further work. Definitely check and mention what year this starts (can we put dates in the plot?). Gut feeling says it's 2008. The term phase transition might be a bit too loaded, not much is lost if we avoid it. I don't understand how the fact that the hockey stick is invariant contradicts the fact that there is a change in the correlation structure. They are things different things enough that one does not necessarily bear any causality on the other (or I'd ask you to argue why not?).}

%\begin{figure}
%\centering
%\begin{subfigure}{0.475\textwidth}
%\includegraphics[width=\textwidth]{Figures/AR_heatmatrix_clusters.eps}
%\caption{\label{fig:AR_heatmatrix_clusters}}
%\end{subfigure}
%\begin{subfigure}{0.475\textwidth}
%\includegraphics[width=\textwidth]{Figures/AR_heatmatrix_CompInd.eps}
%\caption{\label{fig:AR_heatmatrix_CompInd}}
%\end{subfigure}
%\caption{(Left) \subref{fig:AR_heatmatrix_clusters} is a heat map of the ARI computed between the DBHT clusterings of indicators for each pair of $16$ rolling time windows of length $4$, with a shift of $1$ year. (Right) \subref{fig:AR_heatmatrix_CompInd} is the heat map of the ARI computed between the DBHT clusterings of the CDCIs, using the same parameters for the rolling window as \subref{fig:AR_heatmatrix_clusters}. The colour legend for both plots is on the right.}
%\end{figure}

\section{Conclusion}
In this paper, we have investigated whether the collection of development indicators given by the WDI database can be divided using their fundamental topic description. Leveraging on PCA and a novel application of information filtering and hierarchical clustering techniques, we showed that the structure of the topics does not mirror the actual structure between the indicators. This suggests that composite development indicators that are aggregated from restricted sets may ignore key information. Instead, we propose a new set of cluster driven composite development indicators that overcomes these problems. They are objective, data driven, interpretable and are able to make valid comparisons between countries. We have used the composite indicators and some highly influential PageRank indicators to give new insights into the development of countries. Some of these may support decisions for policy makers. Lastly, we showed that our proposed composite indicators can outperform schemes of indicators based on a random benchmark and PageRank. We mention that it has been pointed out by \cite{Mazziotta2019} that using the correlation matrix to form composite indicators may ignore the presence of causality relationships. We mainly use the CDCIs to group countries to understand how they can be classified in terms of their development so that there is no implicit reliance on there being a different causal relationship between indicators. It would, however, be interesting to develop combining the methodology proposed here with an analysis of the causal relationships in a future work.

\section*{Declarations}

\subsection*{Availability of data and materials}
The WDI dataset analysed during the current study is available at the following link \url{https://datacatalog.worldbank.org/dataset/world-development-indicators}.

\subsection*{Competing interests}
  The authors declare that they have no competing interests.
	
\subsection*{Funding}
A.V wishes to thank EPSRC for providing funding during his PhD studies under grant EP/M50788X/1. We also wish to thank the ESRC Network Plus project ’Rebuilding macroeconomics’. We acknowledge support
from Economic and Physical Science Research Council (EPSRC) grant EP/P031730/1. We are grateful to the NVIDIA corporation for supporting our research in this area with the donation of a GPU.

\subsection*{Author's contributions}
 A.V, O.A and T.D.M conceived the experiment(s),  A.V. and O.A. conducted the experiment(s) and A.V, O.A and T.D.M authors analysed the results.  A.V, O.A and T.D.M reviewed the manuscript.

\subsection*{Acknowledgments}
We acknowledge the input of Giuseppe Brandi in this article. We thank the two anonymous reviewers, whose contributions have greatly improved the manuscript.  
 
%%  After submission of the .TEX file,                     %%
%%  you will be prompted to submit your .BBL file.         %%

% if your bibliography is in bibtex format, use those commands:
\bibliographystyle{naturemag-doi} % Style BST file (bmc-mathphys, vancouver, spbasic).
\bibliography{WDI_paper}      % Bibliography file (usually '*.bib' )

\begin{thebibliography}{10}
\urlstyle{rm}
\expandafter\ifx\csname url\endcsname\relax
  \def\url#1{\texttt{#1}}\fi
\expandafter\ifx\csname urlprefix\endcsname\relax\def\urlprefix{URL }\fi
\expandafter\ifx\csname doiprefix\endcsname\relax\def\doiprefix{DOI: }\fi
\providecommand{\bibinfo}[2]{#2}
\providecommand{\eprint}[2][]{\url{#2}}

\bibitem{Stock1989}
\bibinfo{author}{Stock, J.~H.} \& \bibinfo{author}{Watson, M.~W.}
\newblock \bibinfo{journal}{\bibinfo{title}{New indexes of coincident and
  leading economic indicators}}.
\newblock {\emph{\JournalTitle{NBER macroeconomics annual}}}
  \textbf{\bibinfo{volume}{4}}, \bibinfo{pages}{351--394}
  (\bibinfo{year}{1989}).

\bibitem{Mugge2016}
\bibinfo{author}{M{\"u}gge, D.}
\newblock \bibinfo{journal}{\bibinfo{title}{Studying macroeconomic indicators
  as powerful ideas}}.
\newblock {\emph{\JournalTitle{Journal of European Public Policy}}}
  \textbf{\bibinfo{volume}{23}}, \bibinfo{pages}{410--427}
  (\bibinfo{year}{2016}).

\bibitem{Ricardo1891}
\bibinfo{author}{Ricardo, D.}
\newblock \emph{\bibinfo{title}{Principles of political economy and taxation}}
  (\bibinfo{publisher}{G. Bell}, \bibinfo{address}{London},
  \bibinfo{year}{1891}).

\bibitem{Leontief1956}
\bibinfo{author}{Leontief, W.}
\newblock \bibinfo{journal}{\bibinfo{title}{Factor proportions and the
  structure of american trade: further theoretical and empirical analysis}}.
\newblock {\emph{\JournalTitle{Review of Economics and Statistics}}}
  \textbf{\bibinfo{volume}{38}}, \bibinfo{pages}{386--407}
  (\bibinfo{year}{1956}).

\bibitem{Bowen1986}
\bibinfo{author}{Bowen, H.~P.}, \bibinfo{author}{Leamer, E.~E.} \&
  \bibinfo{author}{Sveikauskas, L.~a.}
\newblock \bibinfo{title}{Multicountry, multifactor tests of the factor
  abundance theory} (\bibinfo{year}{1986}).

\bibitem{Aghion1990}
\bibinfo{author}{Aghion, P.} \& \bibinfo{author}{Howitt, P.}
\newblock \bibinfo{title}{A model of growth through creative destruction}.
\newblock \bibinfo{type}{Tech. Rep.}, \bibinfo{institution}{National Bureau of
  Economic Research} (\bibinfo{year}{1990}).

\bibitem{Heckscher1991}
\bibinfo{author}{Heckscher, E.~F.} \& \bibinfo{author}{Ohlin, B.~G.}
\newblock \emph{\bibinfo{title}{Heckscher-Ohlin trade theory}}
  (\bibinfo{publisher}{The MIT Press}, \bibinfo{address}{Cambridge},
  \bibinfo{year}{1991}).

\bibitem{Kremer1993}
\bibinfo{author}{Kremer, M.}
\newblock \bibinfo{journal}{\bibinfo{title}{The o-ring theory of economic
  development}}.
\newblock {\emph{\JournalTitle{The Quarterly Journal of Economics}}}
  \textbf{\bibinfo{volume}{108}}, \bibinfo{pages}{551--575}
  (\bibinfo{year}{1993}).

\bibitem{Krueger2001}
\bibinfo{author}{Krueger, A.~B.} \& \bibinfo{author}{Lindahl, M.}
\newblock \bibinfo{journal}{\bibinfo{title}{Education for growth: why and for
  whom?}}
\newblock {\emph{\JournalTitle{Journal of economic literature}}}
  \textbf{\bibinfo{volume}{39}}, \bibinfo{pages}{1101--1136}
  (\bibinfo{year}{2001}).

\bibitem{Egert2009}
\bibinfo{author}{Egert, B.}, \bibinfo{author}{Kozluk, T.~J.} \&
  \bibinfo{author}{Sutherland, D.}
\newblock \bibinfo{journal}{\bibinfo{title}{Infrastructure and growth:
  empirical evidence}}.
\newblock {\emph{\JournalTitle{CESifo Working Paper Series}}}
  (\bibinfo{year}{2009}).

\bibitem{Aghion2010}
\bibinfo{author}{Aghion, P.}, \bibinfo{author}{Howitt, P.} \&
  \bibinfo{author}{Murtin, F.}
\newblock \bibinfo{title}{The relationship between health and growth: when
  lucas meets nelson-phelps}.
\newblock \bibinfo{type}{Tech. Rep.}, \bibinfo{institution}{National Bureau of
  Economic Research} (\bibinfo{year}{2010}).

\bibitem{United1997}
\bibinfo{author}{(Ghana), U. N. D.~P.}
\newblock \emph{\bibinfo{title}{Ghana Human Development Report}}
  (\bibinfo{publisher}{United Nations Development Programme},
  \bibinfo{address}{Accra}, \bibinfo{year}{1997}).

\bibitem{Salzman2003}
\bibinfo{author}{Salzman, J.}
\newblock \emph{\bibinfo{title}{Methodological choices encountered in the
  construction of composite indices of economic and social well-being}}
  (\bibinfo{publisher}{Centre for the study of living standards},
  \bibinfo{address}{Ottawa}, \bibinfo{year}{2003}).

\bibitem{Sagar1998}
\bibinfo{author}{Sagar, A.~D.} \& \bibinfo{author}{Najam, A.}
\newblock \bibinfo{journal}{\bibinfo{title}{The human development index: a
  critical review}}.
\newblock {\emph{\JournalTitle{Ecological economics}}}
  \textbf{\bibinfo{volume}{25}}, \bibinfo{pages}{249--264}
  (\bibinfo{year}{1998}).

\bibitem{Todaro2015}
\bibinfo{author}{Todaro, M.~P.} \& \bibinfo{author}{Smith, S.~C.}
\newblock \emph{\bibinfo{title}{Economic development}}
  (\bibinfo{publisher}{Pearson}, \bibinfo{address}{New Jersey},
  \bibinfo{year}{2015}).

\bibitem{GCI}
\bibinfo{author}{Huawei}.
\newblock \bibinfo{title}{Global connectivity index 2018}
  (\bibinfo{year}{2018}).

\bibitem{Bray2012}
\bibinfo{author}{Bray, F.}, \bibinfo{author}{Jemal, A.}, \bibinfo{author}{Grey,
  N.}, \bibinfo{author}{Ferlay, J.} \& \bibinfo{author}{Forman, D.}
\newblock \bibinfo{journal}{\bibinfo{title}{Global cancer transitions according
  to the human development index (2008--2030): a population-based study}}.
\newblock {\emph{\JournalTitle{The lancet oncology}}}
  \textbf{\bibinfo{volume}{13}}, \bibinfo{pages}{790--801}
  (\bibinfo{year}{2012}).

\bibitem{Huggins2003}
\bibinfo{author}{Huggins, R.}
\newblock \bibinfo{journal}{\bibinfo{title}{Creating a uk competitiveness
  index: regional and local benchmarking}}.
\newblock {\emph{\JournalTitle{Regional Studies}}}
  \textbf{\bibinfo{volume}{37}}, \bibinfo{pages}{89--96}
  (\bibinfo{year}{2003}).

\bibitem{Maaten2009}
\bibinfo{author}{Van Der~Maaten, L.}, \bibinfo{author}{Postma, E.} \&
  \bibinfo{author}{Van~den Herik, J.}
\newblock \bibinfo{journal}{\bibinfo{title}{Dimensionality reduction: a
  comparative}}.
\newblock {\emph{\JournalTitle{J Mach Learn Res}}}
  \textbf{\bibinfo{volume}{10}}, \bibinfo{pages}{66--71}
  (\bibinfo{year}{2009}).

\bibitem{Bun2017}
\bibinfo{author}{Bun, J.}, \bibinfo{author}{Bouchaud, J.-P.} \&
  \bibinfo{author}{Potters, M.}
\newblock \bibinfo{journal}{\bibinfo{title}{Cleaning large correlation
  matrices: tools from random matrix theory}}.
\newblock {\emph{\JournalTitle{Physics Reports}}}
  \textbf{\bibinfo{volume}{666}}, \bibinfo{pages}{1--109}
  (\bibinfo{year}{2017}).

\bibitem{Page1999}
\bibinfo{author}{Page, L.}, \bibinfo{author}{Brin, S.},
  \bibinfo{author}{Motwani, R.} \& \bibinfo{author}{Winograd, T.}
\newblock \bibinfo{title}{The pagerank citation ranking: Bringing order to the
  web.}
\newblock \bibinfo{type}{Tech. Rep.}, \bibinfo{institution}{Stanford InfoLab}
  (\bibinfo{year}{1999}).

\bibitem{Cristelli2018}
\bibinfo{author}{Cristelli, M.}, \bibinfo{author}{Tacchella, A.} \&
  \bibinfo{author}{Cader, M.}
\newblock \bibinfo{journal}{\bibinfo{title}{The virtuous interplay of
  infrastructure development and the complexity of nations}}.
\newblock {\emph{\JournalTitle{Entropy}}} \textbf{\bibinfo{volume}{20}},
  \bibinfo{pages}{761} (\bibinfo{year}{2018}).

\bibitem{Lai2003}
\bibinfo{author}{Lai, D.}
\newblock \bibinfo{journal}{\bibinfo{title}{Principal component analysis on
  human development indicators of china}}.
\newblock {\emph{\JournalTitle{Social indicators research}}}
  \textbf{\bibinfo{volume}{61}}, \bibinfo{pages}{319--330}
  (\bibinfo{year}{2003}).

\bibitem{Nardo2005}
\bibinfo{author}{Nardo, M.}, \bibinfo{author}{Saisana, M.},
  \bibinfo{author}{Saltelli, A.} \& \bibinfo{author}{Tarantola, S.}
\newblock \bibinfo{journal}{\bibinfo{title}{Tools for composite indicators
  building}}.
\newblock {\emph{\JournalTitle{European Comission, Ispra}}}
  \textbf{\bibinfo{volume}{15}}, \bibinfo{pages}{19--20}
  (\bibinfo{year}{2005}).

\bibitem{Castellacci2011}
\bibinfo{author}{Castellacci, F.}
\newblock \bibinfo{journal}{\bibinfo{title}{Closing the technology gap?}}
\newblock {\emph{\JournalTitle{Review of Development Economics}}}
  \textbf{\bibinfo{volume}{15}}, \bibinfo{pages}{180--197}
  (\bibinfo{year}{2011}).

\bibitem{Mazziotta2019}
\bibinfo{author}{Mazziotta, M.} \& \bibinfo{author}{Pareto, A.}
\newblock \bibinfo{journal}{\bibinfo{title}{Use and misuse of pca for measuring
  well-being}}.
\newblock {\emph{\JournalTitle{Social Indicators Research}}}
  \textbf{\bibinfo{volume}{142}}, \bibinfo{pages}{451--476}
  (\bibinfo{year}{2019}).

\bibitem{Mantegna1999}
\bibinfo{author}{Mantegna, R.~N.}
\newblock \bibinfo{journal}{\bibinfo{title}{Hierarchical structure in financial
  markets}}.
\newblock {\emph{\JournalTitle{The European Physical Journal B-Condensed Matter
  and Complex Systems}}} \textbf{\bibinfo{volume}{11}},
  \bibinfo{pages}{193--197} (\bibinfo{year}{1999}).

\bibitem{Tumminello2005}
\bibinfo{author}{Tumminello, M.}, \bibinfo{author}{Aste, T.},
  \bibinfo{author}{Di~Matteo, T.} \& \bibinfo{author}{Mantegna, R.~N.}
\newblock \bibinfo{journal}{\bibinfo{title}{A tool for filtering information in
  complex systems}}.
\newblock {\emph{\JournalTitle{Proceedings of the National Academy of
  Sciences}}} \textbf{\bibinfo{volume}{102}}, \bibinfo{pages}{10421--10426}
  (\bibinfo{year}{2005}).

\bibitem{Anderberg2014}
\bibinfo{author}{Anderberg, M.~R.}
\newblock \emph{\bibinfo{title}{Cluster analysis for applications: probability
  and mathematical statistics: a series of monographs and textbooks}},
  vol.~\bibinfo{volume}{19} (\bibinfo{publisher}{Academic press},
  \bibinfo{address}{Cambridge}, \bibinfo{year}{2014}).

\bibitem{Song2012}
\bibinfo{author}{Song, W.-M.}, \bibinfo{author}{Di~Matteo, T.} \&
  \bibinfo{author}{Aste, T.}
\newblock \bibinfo{journal}{\bibinfo{title}{Hierarchical information clustering
  by means of topologically embedded graphs}}.
\newblock {\emph{\JournalTitle{PloS one}}} \textbf{\bibinfo{volume}{7}},
  \bibinfo{pages}{e31929} (\bibinfo{year}{2012}).

\bibitem{Musmeci2015}
\bibinfo{author}{Musmeci, N.}, \bibinfo{author}{Aste, T.} \&
  \bibinfo{author}{Di~Matteo, T.}
\newblock \bibinfo{journal}{\bibinfo{title}{Relation between financial market
  structure and the real economy: comparison between clustering methods}}.
\newblock {\emph{\JournalTitle{PloS one}}} \textbf{\bibinfo{volume}{10}},
  \bibinfo{pages}{e0116201} (\bibinfo{year}{2015}).

\bibitem{Sneath1957}
\bibinfo{author}{Sneath, P.~H.}
\newblock \bibinfo{journal}{\bibinfo{title}{The application of computers to
  taxonomy}}.
\newblock {\emph{\JournalTitle{Microbiology}}} \textbf{\bibinfo{volume}{17}},
  \bibinfo{pages}{201--226} (\bibinfo{year}{1957}).

\bibitem{Graham1985}
\bibinfo{author}{Graham, R.~L.} \& \bibinfo{author}{Hell, P.}
\newblock \bibinfo{journal}{\bibinfo{title}{On the history of the minimum
  spanning tree problem}}.
\newblock {\emph{\JournalTitle{Annals of the History of Computing}}}
  \textbf{\bibinfo{volume}{7}}, \bibinfo{pages}{43--57} (\bibinfo{year}{1985}).

\bibitem{Aste2005}
\bibinfo{author}{Aste, T.}, \bibinfo{author}{Di~Matteo, T.} \&
  \bibinfo{author}{Hyde, S.}
\newblock \bibinfo{journal}{\bibinfo{title}{Complex networks on hyperbolic
  surfaces}}.
\newblock {\emph{\JournalTitle{Physica A: Statistical Mechanics and its
  Applications}}} \textbf{\bibinfo{volume}{346}}, \bibinfo{pages}{20--26}
  (\bibinfo{year}{2005}).

\bibitem{Musmeci2015a}
\bibinfo{author}{Musmeci, N.}, \bibinfo{author}{Aste, T.} \&
  \bibinfo{author}{Di~Matteo, T.}
\newblock \bibinfo{journal}{\bibinfo{title}{Risk diversification: a study of
  persistence with a filtered correlation-network approach}}.
\newblock {\emph{\JournalTitle{Journal of Network Theory in Finance}}}
  \textbf{\bibinfo{volume}{1}}, \bibinfo{pages}{77--98} (\bibinfo{year}{2015}).

\bibitem{World2018}
\bibinfo{author}{Group, W. B. I. E. D. D.~D.}
\newblock \emph{\bibinfo{title}{World development indicators}}
  (\bibinfo{publisher}{World Bank}, \bibinfo{address}{Washington D.C.},
  \bibinfo{year}{2018}).

\bibitem{Jolliffe2002}
\bibinfo{author}{Jolliffe, I.}
\newblock \emph{\bibinfo{title}{Principal component analysis}}
  (\bibinfo{publisher}{Wiley Online Library}, \bibinfo{address}{Hoboken},
  \bibinfo{year}{2002}).

\bibitem{Plerou2002}
\bibinfo{author}{Plerou, V.} \emph{et~al.}
\newblock \bibinfo{journal}{\bibinfo{title}{Random matrix approach to cross
  correlations in financial data}}.
\newblock {\emph{\JournalTitle{Physical Review E}}}
  \textbf{\bibinfo{volume}{65}}, \bibinfo{pages}{066126}
  (\bibinfo{year}{2002}).

\bibitem{Stein2006}
\bibinfo{author}{Stein, S. A.~M.}, \bibinfo{author}{Loccisano, A.~E.},
  \bibinfo{author}{Firestine, S.~M.} \& \bibinfo{author}{Evanseck, J.~D.}
\newblock \bibinfo{journal}{\bibinfo{title}{Principal components analysis: a
  review of its application on molecular dynamics data}}.
\newblock {\emph{\JournalTitle{Annual Reports in Computational Chemistry}}}
  \textbf{\bibinfo{volume}{2}}, \bibinfo{pages}{233--261}
  (\bibinfo{year}{2006}).

\bibitem{Marchenko1967}
\bibinfo{author}{Mar{\v{c}}enko, V.~A.} \& \bibinfo{author}{Pastur, L.~A.}
\newblock \bibinfo{journal}{\bibinfo{title}{Distribution of eigenvalues for
  some sets of random matrices}}.
\newblock {\emph{\JournalTitle{Sbornik: Mathematics}}}
  \textbf{\bibinfo{volume}{1}}, \bibinfo{pages}{457--483}
  (\bibinfo{year}{1967}).

\bibitem{Mishra2008}
\bibinfo{author}{Mishra, S.~K.}
\newblock \bibinfo{journal}{\bibinfo{title}{On construction of robust composite
  indices by linear aggregation}}.
\newblock {\emph{\JournalTitle{Available at SSRN 1147964}}}
  (\bibinfo{year}{2008}).

\bibitem{Bishop2006}
\bibinfo{author}{Bishop, C.~M.}
\newblock \emph{\bibinfo{title}{Pattern recognition and machine learning}}
  (\bibinfo{publisher}{springer}, \bibinfo{address}{Berlin},
  \bibinfo{year}{2006}).

\bibitem{Ravasz2003}
\bibinfo{author}{Ravasz, E.} \& \bibinfo{author}{Barab{\'a}si, A.-L.}
\newblock \bibinfo{journal}{\bibinfo{title}{Hierarchical organization in
  complex networks}}.
\newblock {\emph{\JournalTitle{Physical review E}}}
  \textbf{\bibinfo{volume}{67}}, \bibinfo{pages}{026112}
  (\bibinfo{year}{2003}).

\bibitem{Corominas2013}
\bibinfo{author}{Corominas-Murtra, B.}, \bibinfo{author}{Go{\~n}i, J.},
  \bibinfo{author}{Sol{\'e}, R.~V.} \& \bibinfo{author}{Rodr{\'\i}guez-Caso,
  C.}
\newblock \bibinfo{journal}{\bibinfo{title}{On the origins of hierarchy in
  complex networks}}.
\newblock {\emph{\JournalTitle{Proceedings of the National Academy of
  Sciences}}} \textbf{\bibinfo{volume}{110}}, \bibinfo{pages}{13316--13321}
  (\bibinfo{year}{2013}).

\bibitem{Jain1999}
\bibinfo{author}{Jain, A.~K.}, \bibinfo{author}{Murty, M.~N.} \&
  \bibinfo{author}{Flynn, P.~J.}
\newblock \bibinfo{journal}{\bibinfo{title}{Data clustering: a review}}.
\newblock {\emph{\JournalTitle{ACM computing surveys (CSUR)}}}
  \textbf{\bibinfo{volume}{31}}, \bibinfo{pages}{264--323}
  (\bibinfo{year}{1999}).

\bibitem{Wang2002}
\bibinfo{author}{Wang, H.}, \bibinfo{author}{Wang, W.}, \bibinfo{author}{Yang,
  J.} \& \bibinfo{author}{Yu, P.~S.}
\newblock \bibinfo{title}{Clustering by pattern similarity in large data sets}.
\newblock In \emph{\bibinfo{booktitle}{Proceedings of the 2002 ACM SIGMOD
  international conference on Management of data}}, \bibinfo{pages}{394--405}
  (\bibinfo{year}{2002}).

\bibitem{Mantegna2000}
\bibinfo{author}{Mantegna, R.~N.} \& \bibinfo{author}{Stanley, H.~E.}
\newblock \emph{\bibinfo{title}{Introduction to econophysics: correlations and
  complexity in finance}} (\bibinfo{publisher}{Cambridge university press},
  \bibinfo{year}{1999}).

\bibitem{Winkler2011}
\bibinfo{author}{Winkler, H.} \emph{et~al.}
\newblock \bibinfo{journal}{\bibinfo{title}{Access and affordability of
  electricity in developing countries}}.
\newblock {\emph{\JournalTitle{World development}}}
  \textbf{\bibinfo{volume}{39}}, \bibinfo{pages}{1037--1050}
  (\bibinfo{year}{2011}).

\bibitem{Garcia2006}
\bibinfo{author}{Garcia-Moreno, C.} \emph{et~al.}
\newblock \bibinfo{journal}{\bibinfo{title}{Prevalence of intimate partner
  violence: findings from the who multi-country study on women's health and
  domestic violence}}.
\newblock {\emph{\JournalTitle{The lancet}}} \textbf{\bibinfo{volume}{368}},
  \bibinfo{pages}{1260--1269} (\bibinfo{year}{2006}).

\bibitem{Smith2000}
\bibinfo{author}{Smith, L.~C.} \& \bibinfo{author}{Haddad, L.~J.}
\newblock \emph{\bibinfo{title}{Explaining child malnutrition in developing
  countries: A cross-country analysis}}, vol. \bibinfo{volume}{111}
  (\bibinfo{publisher}{Intl Food Policy Res Inst}, \bibinfo{address}{Washington
  D.C.}, \bibinfo{year}{2000}).

\bibitem{Ravallion1997}
\bibinfo{author}{Ravallion, M.}
\newblock \bibinfo{journal}{\bibinfo{title}{Can high-inequality developing
  countries escape absolute poverty?}}
\newblock {\emph{\JournalTitle{Economics letters}}}
  \textbf{\bibinfo{volume}{56}}, \bibinfo{pages}{51--57}
  (\bibinfo{year}{1997}).

\bibitem{Bose2007}
\bibinfo{author}{Bose, N.}, \bibinfo{author}{Haque, M.~E.} \&
  \bibinfo{author}{Osborn, D.~R.}
\newblock \bibinfo{journal}{\bibinfo{title}{Public expenditure and economic
  growth: A disaggregated analysis for developing countries}}.
\newblock {\emph{\JournalTitle{The Manchester School}}}
  \textbf{\bibinfo{volume}{75}}, \bibinfo{pages}{533--556}
  (\bibinfo{year}{2007}).

\bibitem{Gupta2008}
\bibinfo{author}{Gupta, G.~R.}, \bibinfo{author}{Parkhurst, J.~O.},
  \bibinfo{author}{Ogden, J.~A.}, \bibinfo{author}{Aggleton, P.} \&
  \bibinfo{author}{Mahal, A.}
\newblock \bibinfo{journal}{\bibinfo{title}{Structural approaches to hiv
  prevention}}.
\newblock {\emph{\JournalTitle{The Lancet}}} \textbf{\bibinfo{volume}{372}},
  \bibinfo{pages}{764--775} (\bibinfo{year}{2008}).

\bibitem{Montgomery2007}
\bibinfo{author}{Montgomery, M.~A.} \& \bibinfo{author}{Elimelech, M.}
\newblock \bibinfo{title}{Water and sanitation in developing countries:
  including health in the equation} (\bibinfo{year}{2007}).

\bibitem{Romer1990}
\bibinfo{author}{Romer, P.~M.}
\newblock \bibinfo{journal}{\bibinfo{title}{Endogenous technological change}}.
\newblock {\emph{\JournalTitle{Journal of political Economy}}}
  \textbf{\bibinfo{volume}{98}}, \bibinfo{pages}{S71--S102}
  (\bibinfo{year}{1990}).

\bibitem{Rand1971}
\bibinfo{author}{Rand, W.~M.}
\newblock \bibinfo{journal}{\bibinfo{title}{Objective criteria for the
  evaluation of clustering methods}}.
\newblock {\emph{\JournalTitle{Journal of the American Statistical
  association}}} \textbf{\bibinfo{volume}{66}}, \bibinfo{pages}{846--850}
  (\bibinfo{year}{1971}).

\bibitem{Tumminello2011}
\bibinfo{author}{Tumminello, M.}, \bibinfo{author}{Micciche, S.},
  \bibinfo{author}{Lillo, F.}, \bibinfo{author}{Piilo, J.} \&
  \bibinfo{author}{Mantegna, R.~N.}
\newblock \bibinfo{journal}{\bibinfo{title}{Statistically validated networks in
  bipartite complex systems}}.
\newblock {\emph{\JournalTitle{PloS one}}} \textbf{\bibinfo{volume}{6}},
  \bibinfo{pages}{e17994} (\bibinfo{year}{2011}).

\bibitem{Feller2008}
\bibinfo{author}{Feller, W.}
\newblock \emph{\bibinfo{title}{An introduction to probability theory and its
  applications}}, vol.~\bibinfo{volume}{2} (\bibinfo{publisher}{John Wiley \&
  Sons}, \bibinfo{address}{Hoboken}, \bibinfo{year}{2008}).

\bibitem{Somarriba2009}
\bibinfo{author}{Somarriba, N.} \& \bibinfo{author}{Pena, B.}
\newblock \bibinfo{journal}{\bibinfo{title}{Synthetic indicators of quality of
  life in europe}}.
\newblock {\emph{\JournalTitle{Social Indicators Research}}}
  \textbf{\bibinfo{volume}{94}}, \bibinfo{pages}{115--133}
  (\bibinfo{year}{2009}).

\bibitem{Pugliese2017}
\bibinfo{author}{Pugliese, E.}, \bibinfo{author}{Chiarotti, G.~L.},
  \bibinfo{author}{Zaccaria, A.} \& \bibinfo{author}{Pietronero, L.}
\newblock \bibinfo{journal}{\bibinfo{title}{Complex economies have a lateral
  escape from the poverty trap}}.
\newblock {\emph{\JournalTitle{PloS one}}} \textbf{\bibinfo{volume}{12}},
  \bibinfo{pages}{e0168540} (\bibinfo{year}{2017}).

\bibitem{Keller2006}
\bibinfo{author}{Keller, K.~R.}
\newblock \bibinfo{journal}{\bibinfo{title}{Investment in primary, secondary,
  and higher education and the effects on economic growth}}.
\newblock {\emph{\JournalTitle{Contemporary Economic Policy}}}
  \textbf{\bibinfo{volume}{24}}, \bibinfo{pages}{18--34}
  (\bibinfo{year}{2006}).

\bibitem{Ross1999}
\bibinfo{author}{Ross, M.~L.}
\newblock \bibinfo{journal}{\bibinfo{title}{The political economy of the
  resource curse}}.
\newblock {\emph{\JournalTitle{World politics}}} \textbf{\bibinfo{volume}{51}},
  \bibinfo{pages}{297--322} (\bibinfo{year}{1999}).

\bibitem{Liu2005}
\bibinfo{author}{Liu, X.}, \bibinfo{author}{Bollen, J.},
  \bibinfo{author}{Nelson, M.~L.} \& \bibinfo{author}{Van~de Sompel, H.}
\newblock \bibinfo{journal}{\bibinfo{title}{Co-authorship networks in the
  digital library research community}}.
\newblock {\emph{\JournalTitle{Information processing \& management}}}
  \textbf{\bibinfo{volume}{41}}, \bibinfo{pages}{1462--1480}
  (\bibinfo{year}{2005}).

\bibitem{Lepenies2016}
\bibinfo{author}{Lepenies, P.}
\newblock \emph{\bibinfo{title}{The power of a single number: a political
  history of GDP}} (\bibinfo{publisher}{Columbia University Press},
  \bibinfo{address}{New York}, \bibinfo{year}{2016}).

\bibitem{Checherita2012}
\bibinfo{author}{Checherita-Westphal, C.} \& \bibinfo{author}{Rother, P.}
\newblock \bibinfo{journal}{\bibinfo{title}{The impact of high government debt
  on economic growth and its channels: An empirical investigation for the euro
  area}}.
\newblock {\emph{\JournalTitle{European economic review}}}
  \textbf{\bibinfo{volume}{56}}, \bibinfo{pages}{1392--1405}
  (\bibinfo{year}{2012}).

\bibitem{Blumenstock2010}
\bibinfo{author}{Blumenstock, J.}, \bibinfo{author}{Shen, Y.} \&
  \bibinfo{author}{Eagle, N.}
\newblock \bibinfo{title}{A method for estimating the relationship between
  phone use and wealth}.
\newblock In \emph{\bibinfo{booktitle}{QualMeetsQuant Workshop at the 4th
  International Conference on Information and Communication Technologies and
  Development}}, vol.~\bibinfo{volume}{13}, \bibinfo{pages}{114--125}
  (\bibinfo{year}{2010}).

\bibitem{Blumenstock2012}
\bibinfo{author}{Blumenstock, J.~E.} \& \bibinfo{author}{Eagle, N.}
\newblock \bibinfo{journal}{\bibinfo{title}{Divided we call: disparities in
  access and use of mobile phones in rwanda}}.
\newblock {\emph{\JournalTitle{Information Technologies \& International
  Development}}} \textbf{\bibinfo{volume}{8}}, \bibinfo{pages}{pp--1}
  (\bibinfo{year}{2012}).

\bibitem{Mehrotra2012}
\bibinfo{author}{Mehrotra, A.}, \bibinfo{author}{Nguyen, A.},
  \bibinfo{author}{Blumenstock, J.} \& \bibinfo{author}{Mohan, V.}
\newblock \bibinfo{title}{Differences in phone use between men and women:
  quantitative evidence from rwanda}.
\newblock In \emph{\bibinfo{booktitle}{Proceedings of the fifth international
  conference on information and communication technologies and development}},
  \bibinfo{pages}{297--306} (\bibinfo{organization}{ACM},
  \bibinfo{year}{2012}).

\bibitem{Gutierrez2013}
\bibinfo{author}{Gutierrez, T.}, \bibinfo{author}{Krings, G.} \&
  \bibinfo{author}{Blondel, V.~D.}
\newblock \bibinfo{journal}{\bibinfo{title}{Evaluating socio-economic state of
  a country analyzing airtime credit and mobile phone datasets}}.
\newblock {\emph{\JournalTitle{arXiv preprint arXiv:1309.4496}}}
  (\bibinfo{year}{2013}).

\bibitem{Gao2019}
\bibinfo{author}{Gao, J.}, \bibinfo{author}{Zhang, Y.-C.} \&
  \bibinfo{author}{Zhou, T.}
\newblock \bibinfo{journal}{\bibinfo{title}{Computational socioeconomics}}.
\newblock {\emph{\JournalTitle{arXiv preprint arXiv:1905.06166}}}
  (\bibinfo{year}{2019}).

\bibitem{Smith2013}
\bibinfo{author}{Smith, C.}, \bibinfo{author}{Mashhadi, A.} \&
  \bibinfo{author}{Capra, L.}
\newblock \bibinfo{journal}{\bibinfo{title}{Ubiquitous sensing for mapping
  poverty in developing countries}}.
\newblock {\emph{\JournalTitle{Paper submitted to the Orange D4D Challenge}}}
  (\bibinfo{year}{2013}).

\bibitem{Thompson2004}
\bibinfo{author}{Thompson, B.}
\newblock \emph{\bibinfo{title}{Exploratory and confirmatory factor analysis:
  Understanding concepts and applications.}} (\bibinfo{publisher}{American
  Psychological Association}, \bibinfo{address}{Washington D.C.},
  \bibinfo{year}{2004}).

\bibitem{fancyimpute}
\bibinfo{author}{Rubinsteyn, A.}, \bibinfo{author}{Feldman, S.},
  \bibinfo{author}{O'Donnell, T.} \& \bibinfo{author}{Beaulieu-Jones, B.}
\newblock \bibinfo{title}{hammerlab/fancyimpute: Version 0.2.0},
  \doiprefix\url{10.5281/zenodo.886614} (\bibinfo{year}{2017}).

\bibitem{mazumder2010spectral}
\bibinfo{author}{Mazumder, R.}, \bibinfo{author}{Hastie, T.} \&
  \bibinfo{author}{Tibshirani, R.}
\newblock \bibinfo{journal}{\bibinfo{title}{Spectral regularization algorithms
  for learning large incomplete matrices}}.
\newblock {\emph{\JournalTitle{Journal of machine learning research}}}
  \textbf{\bibinfo{volume}{11}}, \bibinfo{pages}{2287--2322}
  (\bibinfo{year}{2010}).

\bibitem{troyanskaya2001missing}
\bibinfo{author}{Troyanskaya, O.} \emph{et~al.}
\newblock \bibinfo{journal}{\bibinfo{title}{Missing value estimation methods
  for dna microarrays}}.
\newblock {\emph{\JournalTitle{Bioinformatics}}} \textbf{\bibinfo{volume}{17}},
  \bibinfo{pages}{520--525} (\bibinfo{year}{2001}).

\bibitem{hastie2005elements}
\bibinfo{author}{Hastie, T.}, \bibinfo{author}{Tibshirani, R.},
  \bibinfo{author}{Friedman, J.} \& \bibinfo{author}{Franklin, J.}
\newblock \bibinfo{journal}{\bibinfo{title}{The elements of statistical
  learning: data mining, inference and prediction}}.
\newblock {\emph{\JournalTitle{The Mathematical Intelligencer}}}
  \textbf{\bibinfo{volume}{27}}, \bibinfo{pages}{83--85}
  (\bibinfo{year}{2005}).

\bibitem{tacchella2018dynamical}
\bibinfo{author}{Tacchella, A.}, \bibinfo{author}{Mazzilli, D.} \&
  \bibinfo{author}{Pietronero, L.}
\newblock \bibinfo{journal}{\bibinfo{title}{A dynamical systems approach to
  gross domestic product forecasting}}.
\newblock {\emph{\JournalTitle{Nature Physics}}} \textbf{\bibinfo{volume}{14}},
  \bibinfo{pages}{861} (\bibinfo{year}{2018}).

\bibitem{webber2012bi}
\bibinfo{author}{Webber, J. B.~W.}
\newblock \bibinfo{journal}{\bibinfo{title}{A bi-symmetric log transformation
  for wide-range data}}.
\newblock {\emph{\JournalTitle{Measurement Science and Technology}}}
  \textbf{\bibinfo{volume}{24}}, \bibinfo{pages}{027001}
  (\bibinfo{year}{2012}).

\bibitem{Verma2019}
\bibinfo{author}{Verma, A.}, \bibinfo{author}{Vivo, P.} \&
  \bibinfo{author}{Di~Matteo, T.}
\newblock \bibinfo{journal}{\bibinfo{title}{A memory-based method to select the
  number of relevant components in principal component analysis}}.
\newblock {\emph{\JournalTitle{Journal of Statistical Mechanics: Theory and
  Experiment}}} \textbf{\bibinfo{volume}{2019}}, \bibinfo{pages}{093408}
  (\bibinfo{year}{2019}).

\bibitem{Guhr2003}
\bibinfo{author}{Guhr, T.} \& \bibinfo{author}{K{\"a}lber, B.}
\newblock \bibinfo{journal}{\bibinfo{title}{A new method to estimate the noise
  in financial correlation matrices}}.
\newblock {\emph{\JournalTitle{Journal of Physics A: Mathematical and
  General}}} \textbf{\bibinfo{volume}{36}}, \bibinfo{pages}{3009}
  (\bibinfo{year}{2003}).

\bibitem{Livan2011}
\bibinfo{author}{Livan, G.}, \bibinfo{author}{Alfarano, S.} \&
  \bibinfo{author}{Scalas, E.}
\newblock \bibinfo{journal}{\bibinfo{title}{Fine structure of spectral
  properties for random correlation matrices: An application to financial
  markets}}.
\newblock {\emph{\JournalTitle{Physical Review E}}}
  \textbf{\bibinfo{volume}{84}}, \bibinfo{pages}{016113}
  (\bibinfo{year}{2011}).

\bibitem{Wilinski2018}
\bibinfo{author}{Wilinski, M.}, \bibinfo{author}{Ikeda, Y.} \&
  \bibinfo{author}{Aoyama, H.}
\newblock \bibinfo{journal}{\bibinfo{title}{Complex correlation approach for
  high frequency financial data}}.
\newblock {\emph{\JournalTitle{Journal of Statistical Mechanics: Theory and
  Experiment}}} \textbf{\bibinfo{volume}{2018}}, \bibinfo{pages}{023405}
  (\bibinfo{year}{2018}).

\bibitem{Van2000}
\bibinfo{author}{Van~der Vaart, A.~W.}
\newblock \emph{\bibinfo{title}{Asymptotic statistics}},
  vol.~\bibinfo{volume}{3} (\bibinfo{publisher}{Cambridge university press},
  \bibinfo{year}{2000}).

\bibitem{Zou2005}
\bibinfo{author}{Zou, H.} \& \bibinfo{author}{Hastie, T.}
\newblock \bibinfo{journal}{\bibinfo{title}{Regularization and variable
  selection via the elastic net}}.
\newblock {\emph{\JournalTitle{Journal of the Royal Statistical Society: Series
  B (Statistical Methodology)}}} \textbf{\bibinfo{volume}{67}},
  \bibinfo{pages}{301--320} (\bibinfo{year}{2005}).

\end{thebibliography}
% for author-year bibliography (bmc-mathphys or spbasic)
% a) write to bib file (bmc-mathphys only)
% @settings{label, options="nameyear"}

% or include bibliography directly:
% \begin{thebibliography}
% \bibitem{b1}
% \end{thebibliography}

%%%%%%%%%%%%%%%%%%%%%%%%%%%%%%%%%%%
%%                               %%
%% Figures                       %%
%%                               %%
%% NB: this is for captions and  %%
%% Titles. All graphics must be  %%
%% submitted separately and NOT  %%
%% included in the Tex document  %%
%%                               %%
%%%%%%%%%%%%%%%%%%%%%%%%%%%%%%%%%%%

%%
%% Do not use \listoffigures as most will included as separate files

%\begin{figure}[h!]
%\centering
%\includegraphics[width=\textwidth]{OverExpression.pdf}
%\caption{\csentence{Overexpression of topics in each DBHT cluster.} The number of overexpressed topics, where a topic is overexpressed when the chance of the number of indicators of that topic being present in cluster $k$ (for more details see \cref{DBHT_similarity}) is below a p-value of $8.17\times 10^{-6}$.}
%\label{fig:OverExpression}
%\end{figure}

%%%%%%%%%%%%%%%%%%%%%%%%%%%%%%%%%%%
%%                               %%
%% Tables                        %%
%%                               %%
%%%%%%%%%%%%%%%%%%%%%%%%%%%%%%%%%%%

%% Use of \listoftables is discouraged.
%%
\clearpage
%%%%%%%%%%%%%%%%%%%%%%%%%%%%%%%%%%%
%%                               %%
%% Additional Files              %%
%%                               %%
%%%%%%%%%%%%%%%%%%%%%%%%%%%%%%%%%%%

\section*{Appendix}

\subsection{Imputation} \label{Imputation}
\begin{figure}
\centering
\includegraphics[width=0.7\textwidth]{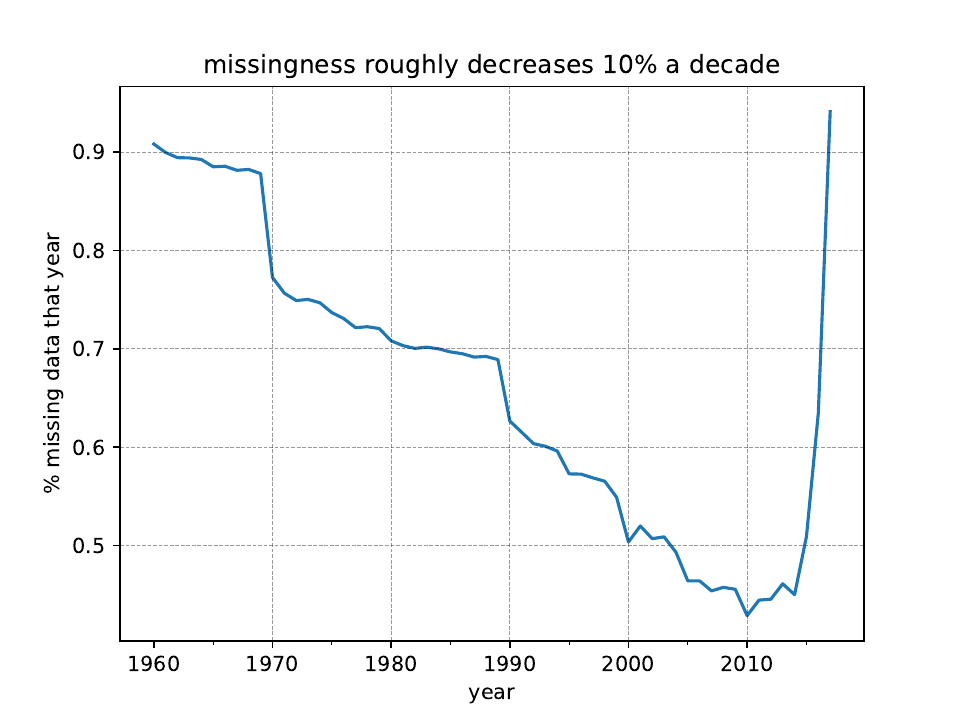}
\caption{Missing data percentage per year, all years (1963-2017), all countries.}
\label{fig:miss_time}
\end{figure}

The WDI dataset suffers from high levels of missing data. We solved this problem with a combination of removal and imputation of datapoints. For starters, the amount of missing data decreases in time, as can be seen in \cref{fig:miss_time}. We decided to use the last 20 years of data, which have the least amount of missing datapoints in the dataset, so to not have to deal with missingness values above 50\%.

% Some countries lack data for almost all indicators and vice-versa some indicators lack data for almost all countries, as can be seen from Figures \cref{fig:miss_c_per_ind} and \cref{fig:miss_ind_per_c}. To mitigate this problem, we removed the worst 225 indicators and the worst 50 countries from the dataset.

\begin{figure}
\centering
\includegraphics[width=0.7\textwidth]{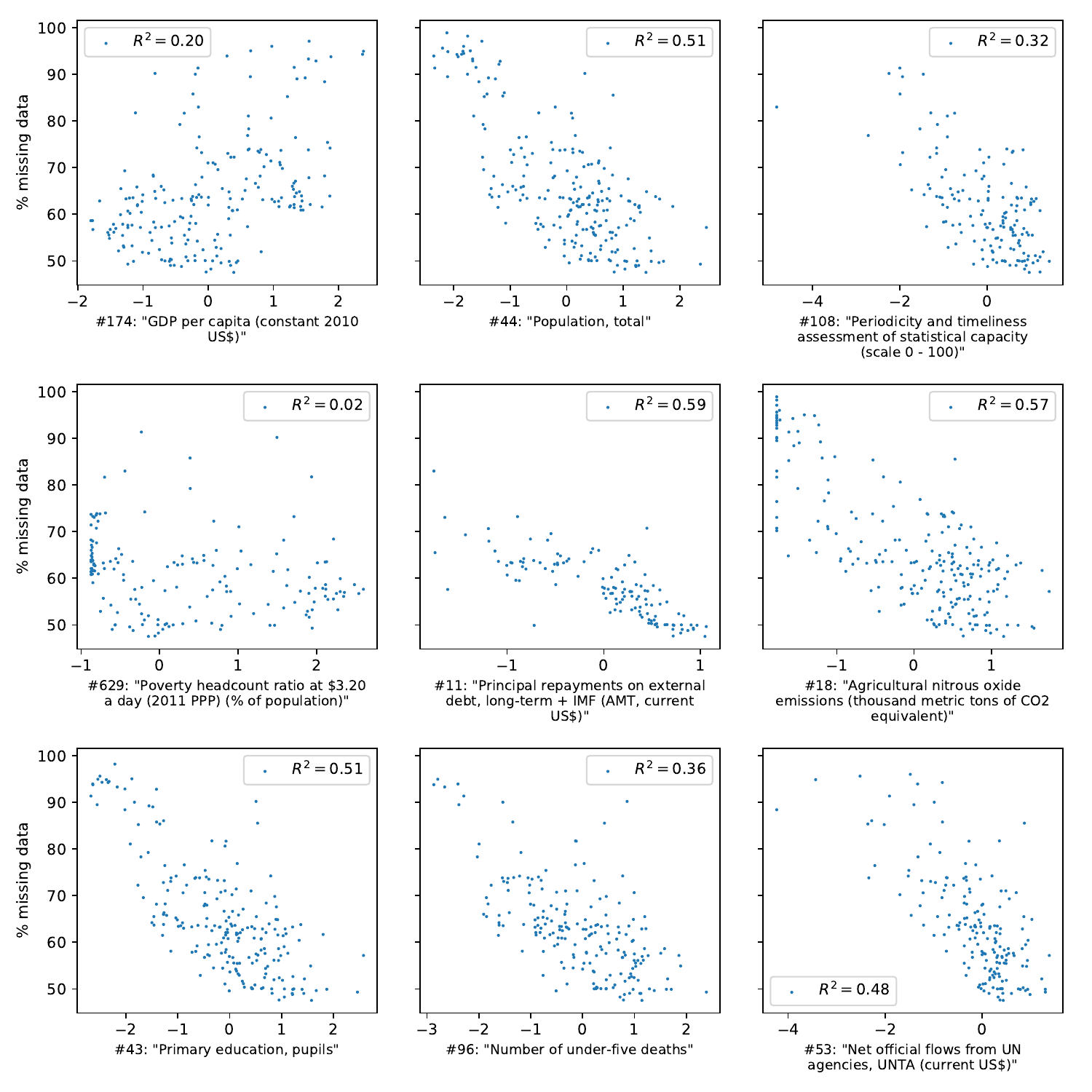}
\caption{Correlation between percentage of missing points for a country and the value of an indicator, all years (1963-2017), all countries.}
\label{fig:miss_factors}
\end{figure}

We considered the possible bias of the dataset due to the fact that data is not missing at random. In fact, it can be seen from \cref{fig:miss_factors} that the amount of missing data a country has is correlated, sometimes strongly, with the values of some of its indicators. It seems that the dataset is biased towards industrialized and more developed countries. While this might cause problems when one tries to make predictions out of the data, we believe the results about the existence of a correlation structure in the data are affected little by this.

%\begin{figure}
%\centering
%\includegraphics[width=0.7\textwidth]{missing_c_per_indicator.pdf}
%\caption{Missing countries per indicator, all years %(1963-2017), all countries.}
%\label{fig:miss_c_per_ind}
%\end{figure}

%\begin{figure}
%\centering
%\includegraphics[width=0.7\textwidth]{missing_ind_per_country.pdf}
%\caption{Missing indicators per country, all years (1963-2017), all countries.}
%\label{fig:miss_ind_per_c}
%\end{figure}

The remaining data still has a high amount of missingness. We therefore proceded to impute it. We tested several algorithms on the dataset, readily available from the Fancyimpute python package \cite{fancyimpute}. They cover mostly matrix factorization approaches to imputation: SoftImpute \cite{mazumder2010spectral}, IterativeSVD \cite{troyanskaya2001missing} and MatrixFactorization \cite{fancyimpute} are all based on this principle. SimpleFill consists in replacing missing entries with the median, and KNN is K-Nearest Neighbours \cite{hastie2005elements}. In \cref{table:impute_compare} we report the Mean Average Error (MAE) and Mean Square Error (MSE) for the techniques adopted (obtained by holding out 0.5\% of the data to test the quality of the results). Interestingly, the best performing technique is K-Nearest Neighbours (KNN). This is in line with the result of \cite{tacchella2018dynamical}, which predicts GDP change over time for a country by averaging the past GDP changes of similar countries, where similarity is measured as an euclidean distance on a space defined by two macroeconomic indicators. This agreement might point to the fact that the most reliable way to model a country is by its similarity to similar countries already observed. The only metaparameter for the KNN algorithm ($D$, the number of neighbours to average) has been chosen by means of grid searching on logarithmically separated values of $D$ and testing on a holdout set of size 0.5\%. \cref{table:impute_meta} shows that the best value for $D$ is either 2 or 4, depending on whether one minimizes MAE or MSE. We chose the average, $D=3$. We have checked that the results do not change qualitatively if $D=2$ or $D=4$ is chosen.
\begin{table}[htbp]
  \centering
  \adjustbox{width=0.5\columnwidth,center}{
    \begin{tabular}{|c|c|c|}
    \hline
           imputer &MAE &MSE\\
    \hline
KNN &0.032636 &0.088496 \\
SoftImpute &0.061094 &0.112322 \\
IterativeSVD &0.112888 &0.181256 \\
MatrixFactorization &0.130820 &0.200824 \\
SimpleFill &0.238268 &0.321349 \\
\hline
\end{tabular}
}
\caption{The Mean Average Error (MAE) and Mean Square Error (MSE) in the second and third columns for the different imputation schemes in the first column.}
\label{table:impute_compare}
\end{table}

\begin{table}[htbp]
  \centering
    \begin{tabular}{|c|c|c|}
    \hline
D &MAE	&MSE\\
	\hline		
1 &0.033483 &0.102603\\
2 &0.030785 &0.091696\\
3 &0.031161 &0.090172\\
4 &0.031588 &0.087745\\
5 &0.032636 &0.088496\\
6 &0.033698 &0.089346\\
8 &0.036122 &0.091546\\
11 &0.039479 &0.094866\\
14 &0.042721 &0.097449\\
18 &0.046539 &0.100678\\
23 &0.050710 &0.104088\\
29 &0.055189 &0.108290\\
37 &0.060167 &0.113266\\
48 &0.065573 &0.119146\\
61 &0.070659 &0.124860\\
78 &0.075911 &0.131326\\
100 &0.081211 &0.137930\\
\hline
\end{tabular}
\caption{Mean Average Error (MAE) (second column) and Mean Squared Error (MSE) (last column) when varying the number of neighbours to average $D$ of the KNN algorithm (first column) using a 0.5\% holdout set size.}
\label{table:impute_meta}
\end{table}

We have investigated the influence of missing data on the results by adding a random white noise, with the value of the variance given by the Mean Square Error (MSE) when $K=3$ (the parameter of the KNN which optimises the MSE). We then recalculated the correlation matrix and reapplied the Directed Bubble Hierarchical Tree (DBHT) algorithm. Comparing the two different set of clusters tests whether the value imputed by the KNN is accurate. Practically, this comparison is achieved by applying a similar procedure that is detailed in section $4.2$ for comparing the DBHT clustering to the topics to see if the clusters in the imputed data are overexpressed. This overexpression indicates statistically whether our imputed data clusters are indeed present in the random data clusters. With a p-value of $1.80\times 10^{-6}$, which is the p-value of $0.01$ modified by Bonferoni correction, that all of the clusters in the imputed data are overexpressed and thus also present into the random data. We also tested whether the KNN algorithm for imputation is appropriate and does not significantly affect the results. This is accomplished by instead replacing the missing data with random white noise with mean $0$ and variance $1$ (since the indicator data are standardised before being imputed). Like before, we recalculate the correlation matrix, apply the DBHT algorithm and compare the two sets of clusters through the same procedure with a p-value of $1.68\times 10^{-6}$. Here, we find $96$ of the original $102$ clusters are overexpressed and thus present in this new random data. This is quite high considering that the assumption tested and total replacement of missing data (rather than just adding noise with a lower standard deviation) here is stronger than before.

Since the formation of the CDCIs relies on the clusters present in the data, we can therefore safely conclude that missing data does not change the final results significantly.

\subsection{Distribution regularisation} \label{Distribution_regularisation}
\begin{figure}
\centering
\includegraphics[width=\textwidth]{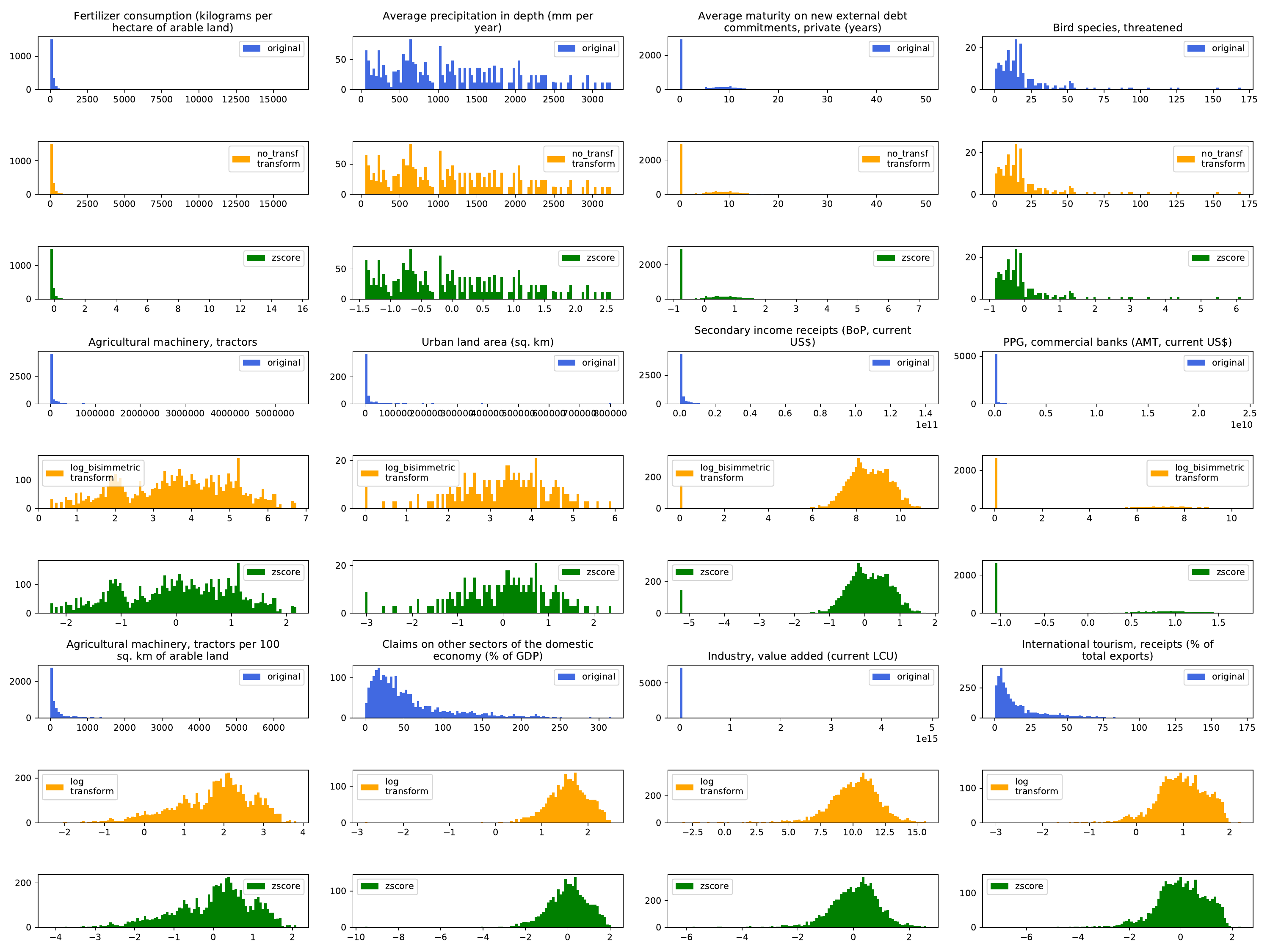}
\caption{Examples of the transformations applied to indicators and how they transform their distribution.}
\label{fig:transf_examples}
\end{figure}

Another characteristic of the WDI dataset is the heterogeneity of the value distributions across different indicators. For example, many indicators are percentages, and as such are bounded between the values 0 and 100. Long-tailed distributions are very common, as well as some that might remind Gaussian distributions. A sample of these distributions can be seen in \cref{fig:transf_examples}. We applied mathematical transformations to some of the indicators, in order to change their distribution and have a more homogeneous and tractable dataset.

We applied one of three possible transformations to each indicator. The first possibility is the identity function, i.e. we left the values unchanged. The second consists in taking the base-10 logarithm of the modulus of each indicator's value. The third is the \emph{bisymmetric log transformation} \cite{webber2012bi}.

\begin{equation}
\text{logbisymmetric}_b(x) = \text{sign}(x) * \log_b(1 + \left|x\right|)
\end{equation}

Given the high number of indicators and the need to avoid arbitrary decisions, the decision of what transformations to apply to each indicator has been made through an algorithm. To understand the criteria used, we will introduce first the definition of \emph{span} of a set of numbers $X$. We define span as:

\begin{equation}
\text{span}(X) = \text{max}_{x \in X}(\log_{10}(|x|)) - \text{min}_{x \in X}(\log_{10}(|x|))
\end{equation}

In order to decide what transformation to apply to each indicator, we consider the set of all values for that indicator found in the dataset, $X$. We then define two quantities. The first we will call \emph{in-span}, which is the span for the subset of values $x$ found in $X$ such that $-1<x<1$. The second is the \emph{out-span}, i.e. the span for all values of X that are outside the $[-1,1]$ interval:

\begin{align*}
\text{inspan}(X) &= \text{span}({x | x \in (X \cap (-1,1)}) \\
\text{outspan}(X) &= \text{span}({x | x \in X \setminus [-1,1])})
\end{align*}

Then, the algorithm for assigning the transformation is this:

\begin{algorithm}[H]
\SetAlgoLined
\KwResult{What kind of transformation to apply to the indicator.}
 given a set of numbers $X$\;
compute $\text{bothsigns}(X)$ = whether $X$ contains both numbers $>0$ and $<0$\;
compute $\text{haszeros}(X)$ = whether $X$ contains the value $0$\;
compute $\text{inspan}(X)$, $\text{outspan}(X)$\;
  \eIf{$\text{outspan}(X) > 2$}{
   \eIf{$\text{inspan}(X) > 2$}{
   	\eIf{not $\text{haszero}(X)$ and not $\text{bothsigns}(X)$}
   		{apply $\log_{10}(|X|)$\;}
   		{apply $\text{identity}(X)$\;}
   	}{apply $\text{logbisymmetric}_{10}(X)$\;}
   }{apply $\text{identity}(X)$\;}
\caption{The algorithm used to choose which transformation to apply to each individual indicator, given the set of values $X$ from its empirical distribution}
\end{algorithm}

%\begin{algorithm}
%given a set of numbers X
%compute both_signs = whether X contains both numbers >0 and <0
%compute has_zeros = whether X contains the value 0
%compute in_span, out_span
%	if out_span > 2:
%		if in_span > 2:
%			if not has_zero and not both_signs:
%				apply \log_{10}(|X|)
%			else:
%				apply \text{identity}(X)
%		else:
%			apply \text{log_bisymmetric}_{10}(X)
%	else:
%		apply \text{identity}(X)
%\end{algorithm}

The rationale behind this algorithm is that frequently the values in X span a large number of orders of magnitude, and in this case we want to transform them so that their distribution is easier to manage with linear techniques such as PCA or factor models. If the numbers are all of the same sign and there is no zero in X, one can directly take the logarithm; otherwise we apply the log-bisymmetric transformation, which has no singularity on the zero and is defined for negative numbers.

After transforming the dataset with this algorithm, we z-score each indicator individually, so to set the mean to zero and the standard deviation to one. A sample of the results of this procedure can be seen in \cref{fig:transf_examples}.

\section{Eigenvalue Spectrum}
\label{EigenvalueSpectrumSupplementary}

Firstly, we should only extract components of $\bm{E}$ that describe relevant interactions between indicators. The question then arises about how many principal components we keep \cite{Jolliffe2002}. This directly controls the size of the reduced correlation matrix - which we would like to be as small as possible - versus the fraction of the total variance of the indicator system that the reduced matrix can explain. It would also help us in identifying what economic indicators are responsible in driving the indicator system by analysing which are the main contributing indicators to the top eigenvalues.

\begin{figure}
\centering
\begin{subfigure}{0.7\textwidth}
\centering
\includegraphics[width=\textwidth]{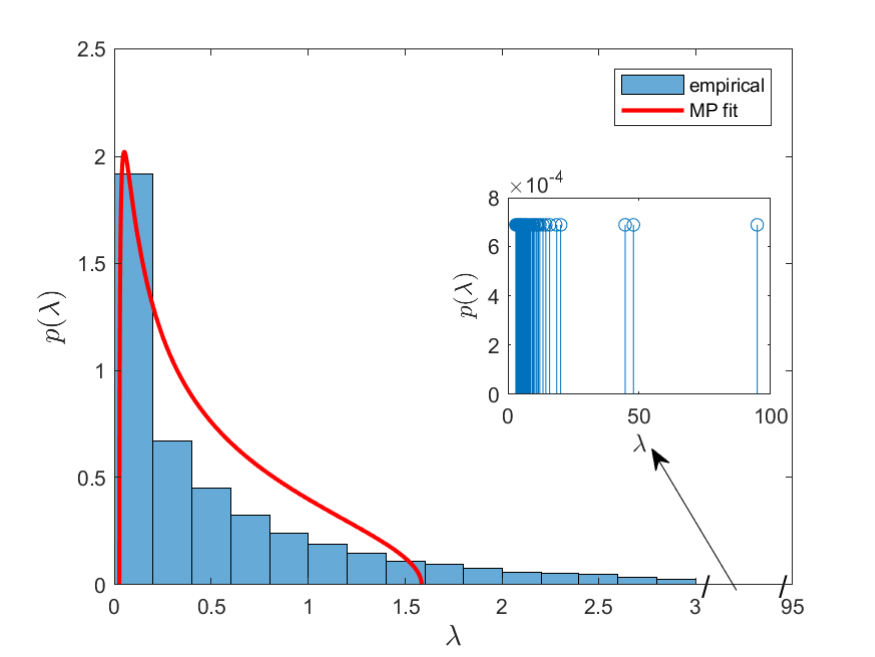}
\caption{Original}
\label{fig:EigenvalueSpectrum}
\end{subfigure}
\\
\begin{subfigure}{0.7\textwidth}
\centering
\includegraphics[width=\textwidth]{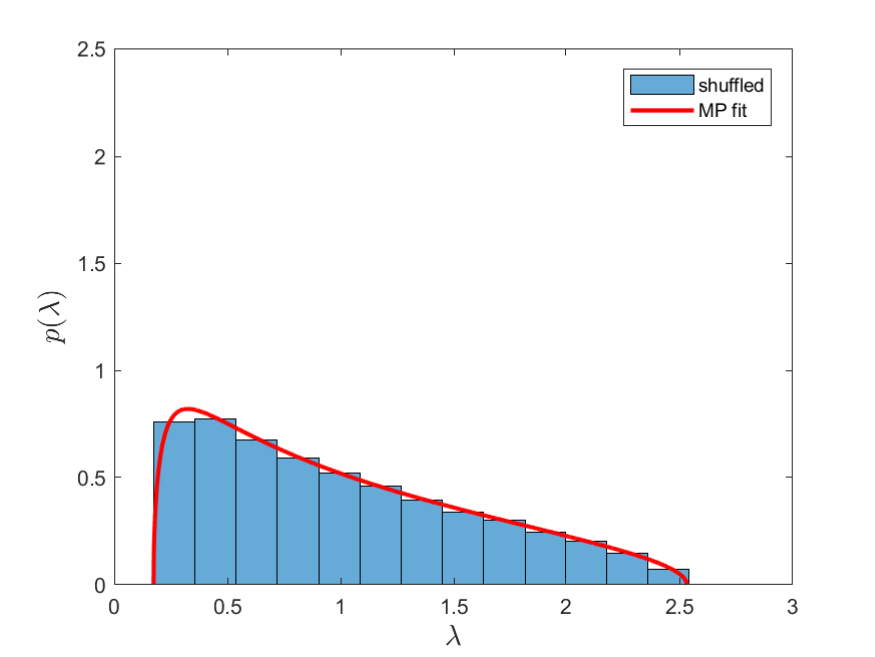}
\caption{Shuffled}
\label{fig:EigenvalueSpectrum_Shuffled}
\end{subfigure}
\caption{(Top) \subref{fig:EigenvalueSpectrum} is the histogram of the empirical eigenvalue distribution for $\mathbf{E}$ in blue bars, with the best MP fit in red. The best MP fit has values $q=0.59\pm 0.025$ and $\sigma=0.71\pm 0.006$. The inset plot are the top $100$ eigenvalues. (Bottom) \subref{fig:EigenvalueSpectrum_Shuffled} is the histogram of the eigenvalue distribution but of the correlation matrix when we shuffle the data. The corresponding MP fit is also given in red, for which $q=0.34\pm 0.027$ and $\sigma=1.00\pm 0.017$.}
\end{figure}

The eigenvalues, however, could also be affected by noise from taking a finite sample \cite{Plerou2002}. We should therefore first study the empirical distribution of the eigenvalues, identifying those eigenvalues which are just noise and discarding them. To identify noisy eigenvalues, we will need a null distribution, produced from a Gaussian white noise process. The answer is provided by the well-known Mar\v{c}enko-Pastur (MP) distribution \cite{Marchenko1967}, given by
\begin{equation}
p(\lambda)=\frac{1}{2\pi q\sigma^{2}}\frac{\sqrt{(\lambda_{+}-\lambda)(\lambda-\lambda_{-})}}{\lambda} \label{MPDist} \ ,
\end{equation}   
where $p(\lambda)$ is the probability density of eigenvalues having support in $\lambda_{-}< \lambda < \lambda_{+}$. The edge points $\lambda_{\pm}=\sigma\left(1\pm \sqrt{q}\right)^{2}$, $q=N/Y$ and $\sigma$ is the standard deviation over all indicators. If we compare the distribution in \cref{MPDist} to the empirical eigenvalue distribution of $\mathbf{E}$, we will be able to see how many components are indistinguishable from noise, often called the 'bulk' eigenvalues. These are then discarded. In practice, this is achieved by fitting Eq. \cref{MPDist} to the eigenvalues of $\mathbf{E}$, with $q$ and $\sigma$ acting as free parameters. The results are shown in \cref{fig:EigenvalueSpectrum} which compares the empirical histogram of eigenvalues of $\mathbf{E}$ and the best fit MP distribution in red, giving $216$
components beyond the upper limit of the MP distribution. Whilst this number appears large, it still means that we can reduce the size of the correlation matrix by $85\%$ before we start to include components which statistically can be seen as noise. Further methods can be used to reduce the number of components further e.g. cross validation or cumulative variance \cite{Jolliffe2002}, and also \cite{Verma2019}.

However, by comparing the best fit MP distribution for our dataset in red in \cref{fig:EigenvalueSpectrum} we see that in fact there seems to be a noticeable deviation of the bulk eigenvalues from the MP distribution, so we can infer that the MP distribution may not be suitable in identifying noisy eigenvalues. We also notice that the best value of $q$ is noticeably different than the theoretical value for this dataset of $0.35$ indicating a significant difference in the predicted properties of the bulk using \cref{MPDist}. Indeed, the use of the MP distribution in this respect has been questioned more recently \cite{Guhr2003,Livan2011,Wilinski2018} for financial data at least.
Moreover, it would also indicate that there could actually be some structure hidden within the bulk eigenvalues. We test this by shuffling our differenced indicator data, recalculating the correlation matrix and again finding the best MP fit to the eigenvalue distribution of this new correlation matrix. In doing so, we destroy the correlations between indicators, therefore testing whether these are the cause of the differences seen in the bulk in \cref{fig:EigenvalueSpectrum}. The results are reported in \cref{fig:EigenvalueSpectrum_Shuffled}, where the histogram of the eigenvalues coming from the new correlation matrix is in blue bars and the best MP fit for this given in red. We can clearly see an almost perfect fit in this case of the MP fit and $q$ much closer to theoretical value which \cref{MPDist} predicts, which suggests that indeed the earlier bulk eigenvalues are a result of non-trivial strucutre within $\mathbf{E}$, and are not just random fluctuations in the data. Overall, these two results together suggest that there is no natural way to a select a subset of principal components without loosing non-trivial information, which may make PCA an unsuitable method of dimensionality reduction for this dataset.

Nevertheless, as the inset plot in \cref{fig:EigenvalueSpectrum} shows, there are some eigenvalues whose magnitude is $2$ times greater than that of some of the smaller eigenvalues e.g. the first principal component has an eigenvalue of $94$. These eigenvalues from the perspective of PCA are the most important eigenvalues since they make the biggest contributions to the overall variance of the system. They are also well separated from the bulk, which means that they are less affected by noise and will have a clearer, more discernible interpretation \cite{Bun2017}.   

\section{Procedure for calculating the p-values of $\rho_{g}$} \label{RhoG_StatTest}
Here we detail the procedure used to calculate the p-values used to produce Table $1$. Under the null hypothesis that $\bm{\rho}_{i}$ is random, the entries of $\Delta \bm{X}$ will be i.i.d normally distributed with mean $0$ and standard deviation of $1$. We can therefore use the exact same definition given in Eq. (3) but with a randomly generated $\Delta \bm{X}$ to produce an instance of $\bm{\rho}_{i}$ under the null hypothesis. One can then estimate the empirical cumulative distribution function \cite{Van2000} of each entry $\rho_{i,g}$ by repeating this process many times and aggregating the results with the same $g$. For Table $1$, we repeat the process $1000$ times.     

\section{PMFG network} \label{PMFG_network}
\begin{figure}
\centering
\includegraphics[width=0.7\textwidth]{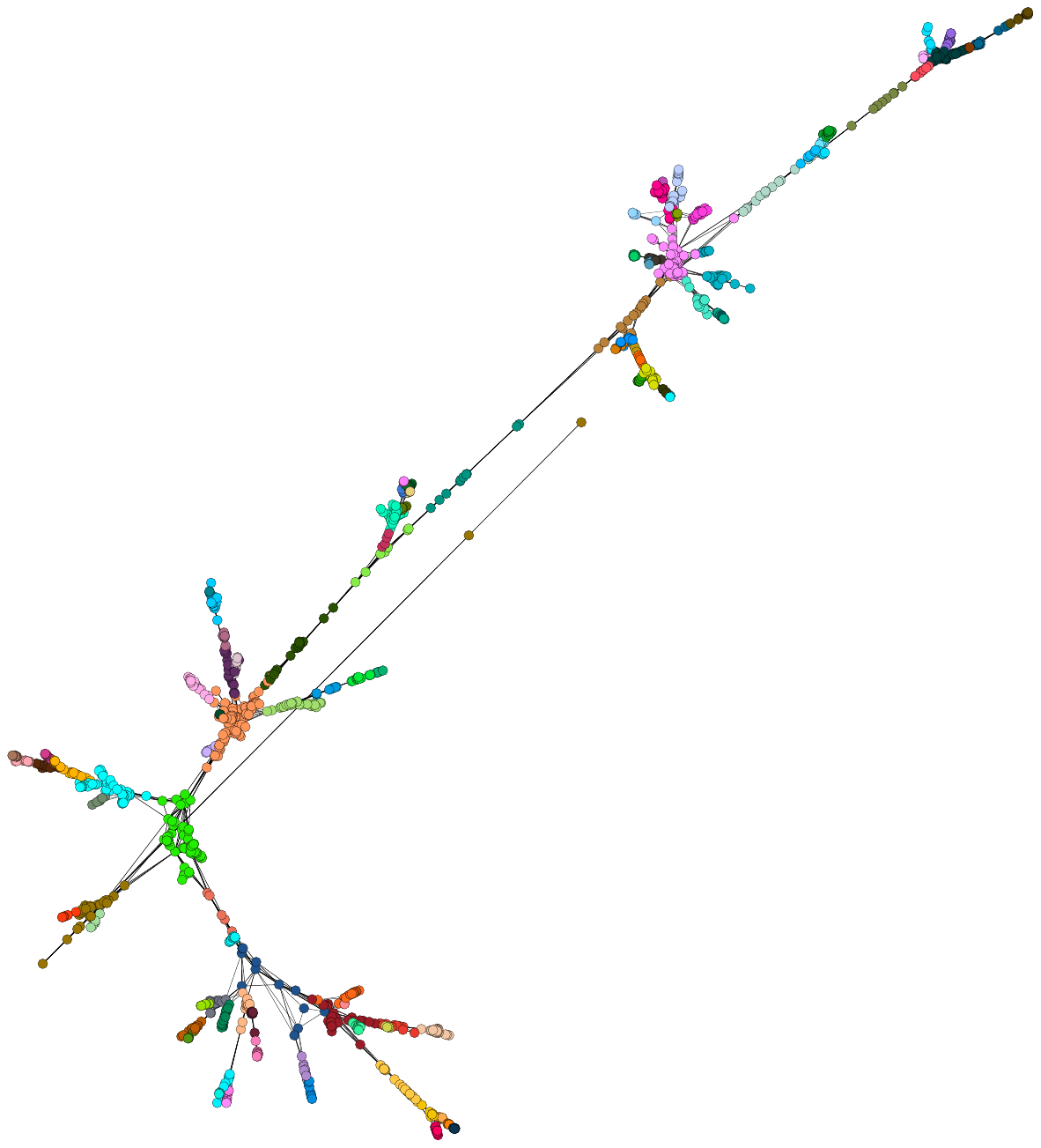}
\caption{The PMFG of $\mathbf{E}$, with the colour of each node representing cluster membership according the DBHT algorithm.}
\label{fig:PMFG_remove}
\end{figure}

Here, we report the visualisation of the PMFG network computed on $\mathbf{E}$ in \cref{fig:PMFG_remove}. From the PMFG, we can observe that there a few hubs of nodes which are connected to other less connected nodes, which is consistent with the observations from other complex networks in different contexts. 

\section{Elastic net regression} \label{ElasticNet}
Elastic net regression is used to find the values of $\beta_{ik}$ from Eq. (5). Further details of the use of this method is provided in this section. Elastic net regression \cite{Zou2005} is a hybrid version of ridge regularisation and lasso regression, thus providing a way of dealing with correlated explanatory variables (in our case $I_{k}(t)$ and $I_{k'}(t)$) and also performing feature selection, which takes into account non-interacting clusters $I_{k'}(t)$ that ridge regularisation would ignore. Elastic net regression solves the constrained minimisation problem
\begin{equation}
\min_{\bm{\beta}_{i}} \frac{1}{Y}\sum_{y=1}^{Y}\left(\Delta\bm{X}(y,i)-\bm{I}^{\dagger}\bm{\beta}_{i}\right)^{2}+\lambda P_{a}(\bm{\beta}_{i}) \ ,
\end{equation}
where $\bm{\beta}_{i}$ is the vector of loadings given by $(\beta_{i1}, \beta_{i2}, \dots,\beta_{iK})^{\dagger}$, $\bm{I}$ is the matrix consisting of columns $(I_{1}(t),I_{2}(t), \dots, I_{K}$ and $\lambda$ and $a$ are hyperparameters. $P_{a}(\bm{\beta}_{i})$ is defined as
\begin{equation}
P_{a}(\bm{\beta}_{i})=\sum_{k=1}^{K}\left((1-a)\frac{\beta_{ik}^{2}}{2}+a |\beta_{ik}|\right) \ . \label{ElasticNetPenalty}
\end{equation}  
The first term in the sum of \cref{ElasticNetPenalty} is the $L_{2}$ penalty for the ridge regularisation and the second term in the sum is the $L_{1}$ penalty for the lasso regression. Hence if $a=0$ then elastic net reduces to ridge regression and if $a=1$ then elastic net becomes lasso, with a value between the two controlling the extent which one is preferred to the other. The determination of the $a$ hyperparameter, controlling the extent of lasso vs ridge, and $\lambda$, for the ridge, is done using 10 cross validated fits \cite{Zou2005}, picking the pair of $(a,\lambda)$ that give the minimum prediction error. 

\end{document}